\documentclass[11pt]{llncs}
\usepackage{amssymb}
\usepackage{xspace}
\usepackage{tikz}
\usepackage{amsmath}
\usepackage{multirow}
\usepackage{scrextend}
\usepackage{wrapfig}
\usepackage{cite}

\newtheorem{algorithm} {Algorithm}
\renewcommand{\iff}{\Leftrightarrow}

\newcommand{\mcomment}[1]{}%\tcp{#1}}
\newcommand{\0}{\ensuremath{\textsf 0}}
\newcommand{\1}{\ensuremath{\textsf 1}}

\newcommand{\A}{\ensuremath{\mathcal{A}}~}

\newcommand{\R}{\ensuremath{\mathcal{R}}\xspace}
\newcommand{\Shield}{\ensuremath{\mathcal{S}}\xspace}

\newcommand{\dcsynth}{DCSYNTH\xspace}

\newcommand {\Or} {\ensuremath{\vee}}
\newcommand {\union} {\ensuremath{\cup}}

\renewcommand {\iff} {\ensuremath{\Leftrightarrow}}

\newcommand {\qddc} {QDDC\xspace}

\newcommand {\ang}[1] {\ensuremath{\langle#1\rangle}}
\newcommand {\sq}[1] {\ensuremath{[#1]}}
\newcommand {\dsq}[1] {\ensuremath{[[#1]]}}
\newcommand {\dcurly}[1] {\ensuremath{\{\{#1\}\}}}

\newcommand {\len}[1] {\ensuremath{len(#1)}}

\newcommand {\intv}[1] {\ensuremath{Intv(#1)}}
\newcommand {\slen} {\ensuremath{slen} }
\newcommand {\scount} {\ensuremath{scount} }
\newcommand {\sdur} {\ensuremath{sdur} }
\newcommand {\pt} {\ensuremath{pt}~}
\newcommand {\ext} {\ensuremath{ext}~}
\newcommand {\true} {\ensuremath{true}~}

\newcommand {\dom}[1] {\ensuremath{dom(#1)}}

\newcommand {\nat} {\ensuremath{\mathbb{N}}}

\newcommand{\oomit}[1]{}

\newcommand{\df}{=}

\newcommand{\DCSYNTH}{DCSYNTH}

\begin{document}
\title{DCSYNTH: Guided Reactive Synthesis with Soft Requirements for Robust Controller and Shield Synthesis}

\author{{Amol Wakankar} \and {Paritosh K. Pandya} \and {Raj Mohan Matteplackel}
}
 
\institute{
Homi Bhabha National Institute, Mumbai, India.\\
Bhabha Atomic Research Centre, Mumbai, India.\\
Email: amolk@barc.gov.in
\and
Tata Institute of Fundamental Research, Mumbai 400005, India.\\
Email: \{pandya, raj.matteplackel\}@tifr.res.in
}

%\author{Amol Wakankar \and Paritosh K.~Pandya \and Raj Mohan Matteplackel }

\maketitle

\setcounter{footnote}{0}

%\IEEEtitleabstractindextext{%

\begin{abstract}
\DCSYNTH\/ is a tool for the synthesis of controllers from safety and bounded liveness requirements given in interval temporal logic \qddc. It investigates the role of soft requirements (with priorities) in obtaining high quality controllers. A \qddc\/ formula specifies past time properties. 
%Its (bounded) counting and regular expression like primitives allow complex quantitative properties to be specified elegantly. 
In \DCSYNTH\/ synthesis, hard requirements must be invariantly satisfied whereas soft requirements may be satisfied "as much as possible" in a best effort manner by the controller. Soft requirements provide an invaluable ability to guide the controller synthesis. In the paper, using \DCSYNTH, 
we show the application of soft requirements in obtaining robust controllers with various specifiable notions of robustness. We also show the use of 
soft requirements to specify and synthesize efficient runtime enforcement shields which can correct burst errors. Finally, we discuss the use of soft requirements in improving the latency of controlled system.
\end{abstract}

%}

\section{Introduction}
\label{section:intro}
A temporal logic formula implicitly specifies the allowed sequence of inputs and outputs.  
In reactive synthesis the aim is to construct a controller (say a Mealy Machine) 
which  explicitly  computes the value of the output sequence 
for any given input sequence, in an online fashion, such that the requirement is met.  
Reactive synthesis is typically a  much harder problem than the monitor synthesis.
Considerable research has been carried out on the reactive synthesis problem %(see \cite{Some survey} for a survey) 
and there are several tools which implement and experiment with reactive synthesis \cite{BBEHJKPRRS15}.

During development systems are often under specified and several different controllers with 
distinct behaviors may all meet the specification. In this case ``guidance'' must be provided 
to the synthesizer to choose amongst them. 
%One criterion can be improved {\bf latencies} in controlled system.
%On the other hand, there may even be contradictory
%requirements which make controller synthesis unrealizable. A solution is to treat some of the
%requirements as ``soft''; they are desirable but not mandatory. 
A critical parameter in acceptance of automatic synthesis technique  is the \emph{quality} of the 
synthesized controller \cite{BEJK14,ABK16}. Thus, just correct-by-construction synthesis is not sufficient.  

Requirements are typically structured as a set of assumptions $A$ and a set of commitments (guarantees) $C$.
Much of the research has addressed ``Be-Correct'' goal \cite{BEJK14} 
for synthesis which states that if assumptions
hold for throughout the behavior then commitment holds for the behavior. 
For safety formulas this would
take the form $G~A \Rightarrow G~C$. 

{\bf Robustness} pertains to the ability of the controller to meet commitments even when (some) environmental assumptions are violated,
and the ability of the controller to recover from transient environmental errors. Laying down such criteria, Bloem et.~al.~have defined 
``don't-be-lazy'' and ``never-give-up'' as desirable synthesis goals \cite{BEJK14}. 
Other criteria include various notions of resilience \cite{ET14}.

A related problem is synthesis of run time enforcement {\bf shield} \cite{BKKW15,WWZ16,ET14,WWYZ17} for critical correctness 
properties. The diagram \ref{fig:safety-shield}
depicts the use of shield, which receives both input $I$ and output $O$ from a (occasionally incorrect) controller. The aim of the shield
is to generate modified output $O'$ which always meets the requirement $Req(I,O')$ even if system output intermittently fails to meet $Req(I,O)$.
Moreover, $O'$ must deviate from $O$ ``as little as possible'' \cite{BKKW15}. 
Bloem et.~al.~proposed a notion $k$-shield for ``as little as possible''
whereas Wu et.~al.~\cite{WWZ16} proposed an alternative notion of ``safety shield'' which tolerates burst errors.

This paper describes a tool \DCSYNTH\/ which allows synthesis of controllers from safety 
and bounded liveness requirements given in 
interval temporal logic {\bf \qddc}.  The paper mainly investigates the role of {\bf soft requirements} (with priorities) in obtaining high quality
controllers.  A \qddc\/ formula specifies past time properties, and it holds at a position in behavior if the past satisfies the property. 
Its (bounded) counting and regular expression like primitives allow complex  quantitative properties to be specified elegantly. 
In \DCSYNTH\/ synthesis, hard requirements must be invariantly satisfied whereas soft requirements may be satisfied 
``as much as possible'' in a best effort manner by the controller. 
In \DCSYNTH\/ specification, the soft requirements (which are \qddc formulas) can be given weights 
and the tool selects from all permissible outputs which meet the hard requirements, 
the one which satisfies a maximal subset of soft requirements, 
in a ``locally optimal fashion''. 
We present the case studies of bus arbiter and 
mine pump specification to illustrate the use of soft requirements in synthesis specification.

Soft requirements provide a powerful and practically useful ability to guide the controller synthesis. 
In the paper, we  show the application of soft requirements in obtaining robust controllers with various 
{\em specifiable} notions of robustness. We also show the use of soft requirements to
specify and synthesize efficient run time enforcement shields which can correct burst errors. 
Finally, we discuss the use of soft requirements in improving the {\bf latency} of controlled system.  
The main contributions of this paper are as follows:
\begin{itemize}
 \item We present a tool \DCSYNTH\/ for synthesis of controllers from \qddc\/ requirements. 
This extends the past work\cite{Pan01a} on model checking interval temporal logic with synthesis abilities. 
 %It is argued elsewhere that \qddc\/ is a powerful and convenient language for specifying safety and bounded liveness properties.
 \item The tool \DCSYNTH\/ allows guided synthesis of controllers based on {\bf soft requirements} which are met ``as much as possible'' in a 
 locally optimal fashion. 
{\em To our knowledge \DCSYNTH\/ is amongst the first reactive synthesis tool to support soft requirement guided synthesis}.
 \item We show application of mixture of hard and soft requirements to {\em specify} various notions of robustness. 
 This includes ``never-give-up'' and ``don't-be-lazy'' criteria of Bloem et.~al.~\cite{BEJK14} as well as 
 $k,b$-resilience criterion of Ehler et.~al.~\cite{ET14}. 
 \DCSYNTH\/ is able to automatically synthesize robust controller from such specification. 
 We give experimental results to evaluate the applicability of our tool for such robust synthesis.
 \item We show application of hard and soft requirements to {\em specify} shields. 
We show how variants of shields of Bloem et.~al.~and Wu et.~al.~can be specified
 and automatically synthesized. 
We give detailed experimental results and comparison with past work to illustrate that \DCSYNTH\/ is able to synthesize
 very compact shields which tolerate burst errors.
 \item Soft goals impact the latency of the controller. 
An associated tool, CTLDC, allows measurement of worst case latencies 
using symbolic techniques for finding longest and shortest paths \cite{Pan05}. 
We give experimental results to show how soft requirements indirectly impact the latencies.
\end{itemize}

The remainder of this paper is organized as follows. Section~\ref{section:motivation} gives a motivating example of synchronous bus arbiter with
hard and soft requirements. Logic \qddc\/ as well as \DCSYNTH\/ syntax are introduced in
Section \ref{section:dcsynth-spec}. 
Section \ref{section:guidedreativesynthesis} gives the guided synthesis method from given \DCSYNTH\/ specification. 
Experimental results in synthesizing controllers using the \DCSYNTH\/ tool are also reported.
Section \ref{section:robustness} deals with robustness and Section \ref{section:shieldsynthesis} 
deals with shield synthesis. Both these include experimental results.
We conclude with Section \ref{section:quantitative} on 
experiments examining impact of soft requirements on controller latencies.
%The paper ends with a brief discussion.
%Section \ref{section:casestudy} gives the experimental results. %We end with a discussion. 
The tool is available for download from \verb|http://www.tcs.tifr.res.in/~pandya/dcsynth/dcsynth.html|. The input files for all the experiments
reported in this paper as well as corresponding outputs of the tool \DCSYNTH\/ are also available there for examination. 

\section{Motivating Example}
\label{section:motivation}
In this section we llustrate the main advantage of guided reactive synthesis 
with \emph{soft requirement} with an example of synchronous bus arbiter.

%\subsection{Synchronous Bus Arbiter}
%\label{sec:arbiter}
An $n$-cell synchronous bus arbiter with $req_i$ as inputs and 
$ack_i$ as corresponding outputs (where $1 \leq i \leq n$), 
is a circuit that arbitrates between a subset of requests 
at each cycle by setting one of the acknowledgments {\em true}. 
Hard requirements include the following three invariant properties.
\begin{equation}
 \begin{array}{l}
  Mutex \df [[ ~\land_{i \neq j} ~\neg ( ack_i \land ack_j) ~]] \\
  NoLoss \df [[ ~~ (\lor_i req_i) \Rightarrow (\lor_j ack_j) ~]] \\
  NoSpurious \df [[ ~~\land_i ~(ack_i \Rightarrow req_i) ~~]] \\
  ARBINV \df Mutex \land NoLoss \land NoSpurious
 \end{array}
\end{equation}
In \qddc\/ $[[ P ]]$ denotes that proposition $P$ is invariantly {\em true}.
Thus, $Mutex$ gives mutual exclusion of acknowledgments. $NoLoss$ states that
if there is at least one request then there must be  an acknowledgment. 
$Nospurious$ states that acknowledgment is only given to a requesting cell. 
%Let their conjunction be called $ARBINV$. 

In literature GR(1) synthesis has been used to specify fairness between  cells of arbiter \cite{BBFJR12}. 
We consider here other variants with concrete bounds on the arbiter response.
\begin{itemize}
\item We can specify response time of $k$ cycles as a bounded liveness property:
let $Resp(req,ack,k)$ denote that if request has been high for last $k$ cycles 
there must have been at least one acknowledgment in the last $k$ cycles 
(next section gives the \qddc formula for this).
%\verb# true^([[req]] && slen=k-1)  => true^(slen=k-1 && <><ack>)#.
Let $ArbResp(n,k)$ state
that for each cell $i$ and for all observation intervals the formula $Resp(req_i,ack_i,k)$ holds. 
\begin{equation}
\begin{array}{l}
  ArbResp(n,k) \df \land_{1 \leq i \leq n} ~[](Resp(req_i,ack_i,k)) \\
  ARBHARD(n,k) \df ARBINV \land ArbResp(n,k)
\end{array}
\end{equation}

\item Consider the following specification with only hard requirements and  no soft requirements,
\begin{equation}
 Arb^{hard}(n,k) \df (ARBHARD, \langle - \rangle) 
\end{equation}
\DCSYNTH\/ can synthesize a controller say $ArbCntrl^{hard}(n,k)$  for given values of $n,k$.
%Table \ref{tab:comparision} gives the results of experiments for various values of $n,k$.
\oomit{
We measure its performance 
based on worst case response time of $i$-th cell.
This is done by computing $MAXLEN(DP,ArbCntrl^{hard}(6,6))$ where $DP$ is $[[ req_i \land \neg ack_i]]$. 
As expected, \DCSYNTH reports that worst case response time for all cells of 
$ArbCntrl^{hard}(6,6)$ is $6$ cycles.
}

\item
Moreover, we can also include {\em soft requirements} giving priority to cells, e.g.
$(ARBHARD(6,6), \langle ack6,ack2 \rangle)$ which gives acknowledgment $ack_6$ as first
preference and $ack_2$ as second preference as far as 
these don't conflict with the hard response requirements. % (cf.~\S\ref{section:casestudy}).
Table.~\ref{tab:comparisionopt} gives experimental results for synthesis with several such soft requirements.

\item If we use  $ARBHARD(6,2)$ in place of $ARBHARD(6,6)$  the specification becomes unrealizable as expected 
(as we cannot guarantee response within two cycles for all 6 cells). 
The tool reports this with a diagnostic counter-strategy.

\item However, we can specify the requirements of response in 2 cycles as 
\emph{soft requirements with priority} as $Arb^{soft}(6,2)$ where
\begin{equation}
 Arb^{soft}(n,k) \df 
 \begin{array}[t]{l}
 (ARBINV, ~~\langle Resp(req_6,ack_6,2),\ldots, \\
 \quad \quad Resp(req_1,ack_1,2) \rangle~)
 \end{array}
\end{equation}
Using \DCSYNTH\/ we get a  controller called $ArbCntrl^{soft}(6,2)$ 
which ``tries'' to give every cell acknowledgment within 2 cycles as far as possible 
with highest priority given to cell 6, followed by cell 5, and so on.
Table.~\ref{tab:comparisionopt} gives the time taken compute this controller. 
Table.~\ref{tab:perfMeasure2}
gives a detailed account of the robust behavior of this  complex arbiter. 
The latency measurement of $ArbCntrl^{soft}(6,2)$ using tool CTLDC shows that the worst case response time for cell 6 is 2 cycles, 
for cell 5 is 3 cycles and for all other cells it is $\infty$. 

\item Consider an $Arb^{hard}(n,k)$ like arbiter working under the assumption 
$Assume(n,i)$ which states that in current cycle at most $i$ requests are 
true  simultaneously. 
Consider the arbiter specification with no soft requirement as follows.

\begin{small}
\begin{equation}
 Arb^{hard}_{assume}(n,k,i) \df ( ( Pref(Assume(n,i)) \Rightarrow ARBHARD(n,k)), \langle - \rangle) 
\label{equation5}
\end{equation}
\end{small}
Synthesis of various  {\em robust} arbiters which function even in presence of {\em intermittent violation} of the
assumption is reported in Table.~\ref{tab:robustArbiterSynthesis}. 

\item Section \ref{section:shieldsynthesis} gives experimental results in specifying and synthesizing run time enforcement shields 
from diverse specifications (see Table.~\ref{tab:dfaBasedComparision}). 
\end{itemize}
Above examples show that \DCSYNTH\/ can fruitfully 
use soft requirements to synthesize better performing, more robust controllers 
as well as shields.

% paritosh
\section{QDDC and DCSynth Specification}
\label{section:dcsynth-spec}

We now give the \qddc formula $Resp(req,ack,k)$ 
for the response time of the arbiter. 
We refer the reader to \S\ref{section:qddc} in Appendix for a discussion on the logic \qddc \cite{Pan01a}. 
%We will start with a formula for mutex.

%\verb# true^([[req]] && slen=k-1)  => true^(slen=k-1 && <><ack>)#.

We define $Resp(req,ack,k)=\verb|(true^([[req]]&&|
\verb|(slen=k-1))=>|\\\verb|(true^(slen=k-1&&(true^<ack>^true)))|$. 
In \qddc the only temporal modality is the {\em chop operator} (\verb|^|) 
and is interpreted over a word $\sigma$ and a closed interval $[i,j]$, 
$0\leq i\leq j<len(\sigma)$, as follows:  
$\sigma,[i,j]\models D_1\textrm{\textasciicircum}D_2$ iff 
$\exists k:i\leq k\leq j:\sigma,[i,k]\models D_1$ and $\sigma,[k,j]\models D_2$. 
For a propositional formula $p$, 
$\sigma,[i,j]\models \textless p\textgreater$ iff $i=j$ and 
$p$ holds at position $i$ in $\sigma$. 
Finally $\sigma, i \models D$ iff $\sigma, [0,i] \models D$.

%Entities such as \slen and \scount are called \emph{terms}. 
The term $\slen=k$ holds for an interval $[i,j]$ if $j-i=k$ and $\scount\ p$, 
where $p$ is a propositional formula, counts 
the number of positions %including the last point 
in the interval where $p$ holds. 
Thus, as {\em true} holds for any word and any interval, 
$Resp$ states that if $req$ holds throughout the last $k$ positions 
%(i.~e.~ $k-1$ cycles, cf.~\S\ref{section:qddc}) 
in the word then at least at one of the last $k$ positions 
$ack$ should hold. 
With $[]$ being {\em for all sub-intervals} operator, 
the formula $ArbResp(n,k)$ says that $Resp(req,ack,k)$ 
must be {\em true} for all intervals and for all the cells from $1$ to $n$.

The tool \dcsynth takes a \emph{\dcsynth spec} as input and outputs a controller. 
%The textual syntax of \emph{DCSynth spec} is exemplified by arbiter specification in Appendix \S\ref{section:arbiter-spec}.
Formally, a {\em \dcsynth spec} is a tuple 
\[
S=(I,O,D^h,\wedge_{i=1}^{i=k}(w_i\iff D_i^s), \langle P_1,\cdots,P_l\rangle)
\]
where $I$ and $O$ are  {\em input} and {\em output} variables of the controller,
respectively. The \qddc formula $D^h$ which is over $I\union O$, 
specifies {\em hard requirement} on the synthesized controller, 
i.~e.~every execution of a controller must satisfy $D^h$ invariantly.
We have a list of indicator definitions $D_1,\ldots,D_k$ where each $D_i^s$ specifies 
a \emph{soft requirement} over $I\union O$. 
% and is associated
%with the {\em indicator variable} $w_i$. 
Each $D_i$ is associated with the {\em indicator variable} $w_i$ which 
witnesses whether $D_i^s$ holds for the execution so far, 
i.~e.~$\sigma, i \models w_i$ iff $\sigma,[0,i] \models D_i^s$.  
Let $W=\{w_i\ |\ 1\leq i\leq k\}$. % be the set of indicator variables. 
The {\em soft requirements} are specified as lexicographically ordered list of propositions 
$ \langle P_1,\cdots,P_l \rangle $ where each $P_i$ is a propositional formula over 
$I\union O\union W$. Each $P_i$ represents a soft requirement with priority higher than 
all $P_j$'s, $1\leq i<j\leq l$. 
%Given two propositional valuations $\nu_1, \nu_2$ over $I \cup O \cup W$,
%we say that $\nu_1 >_{lex} \nu_2$ iff there exist $0\leq i\leq l$ 
%such that $P_i$ is {\em true} under valuation $\nu_1$ and {\em false} under $\nu_2$, 
%and the valuations coincide for all $0\leq j<i:P_j$. 

\begin{example}
A \dcsynth spec for an $n$ cell arbiter with soft requirement 
of $k$ cycle response $ArbResp(n,k)$ for all the cells is as follows: 
let $I=\{req_i\ |\ 1\leq i\leq n\}$ and $O=\{ack_i\ |\ 1\leq i\leq n\}$. 
Let $L=\langle w_6, w_5, \ldots, w_1 \rangle$ which gives the higher numbered cell the higher priority. 
Then 
\[
Arb^{prio}(n,k)=( I,O,ARBINV, ~\wedge_{i=1}^{i=n}(w_i\iff Resp(req_i,ack_i,k)), ~L).
\]
\end{example} 
%The DCSynth spec for 4 cell arbiter (in textual form) 
%is given in Appendix \ref{section:arbiterqddc}, Fig.~\ref{fig:arbitersrc}. 

%Another important aspect of our tool is its ability to specify 
%{\em soft requirements} which can help in synthesizing better controllers. 
%In DCSynth a soft requirement is given as a chain of propositional formulas, 
%in the descending order of priority of the member formulas. 
%The tool, whenever it faces multiple choices, 
%uses the soft requirement 
%and chooses one among the plausible choices which will make 
%the propositional formula with priority as high as possible true. 
%In our experience the soft requirements have done well to engineer better controllers 
%(see Appendix \S\ref{section:} for more one this). 

% raj
\section{Guided Reactive Synthesis Algorithm}
\label{section:guidedreativesynthesis}

Given $S=( I,O,D^h,\wedge_{i=1}^{i=k}(w_i\iff D_i^s), \langle P_1,\cdots,P_l\rangle)$, 
a DCSynth spec, we synthesize a controller as below. %in following steps.

\begin{itemize}
\item The formula $D^{Ind}\df pref(\wedge_{i=1}^{i=k}\verb#(true^<#w_i\verb#><=>#D_i^s\verb#)#)$ 
states that at every point in execution, the value of $w_i$ equals the truth-value of $D_i^s$.
We construct a language equivalent symbolic DFA, called  $A^{Hard+Ind}$, for the formula $D^{hard} \land D^{Ind}$ using tools DCVALID and MONA. 

From ~\cite{Pan01a}, it is known that for every QDDC formula D, 
we can effectively construct a equivalent finite state automaton A(D),
such that a word is accepted by A(D) iff it satisfies formula D.
Tool DCVALID implements this procedure.

\item A {\em safety monitor automaton} $A^{mon}$ is obtained by computing the prefix closure of $A^{Hard+Ind}$. This
automaton has the alphabet $2^{I \cup O \cup W}$. The automaton is reduced to its minimal deterministic form.
This automaton forms the arena on which further synthesis is carried out.
\item The {\em Maximally Permissive Non deterministic Controller} (MPNC) 
is computed from the safety automaton using standard safety synthesis algorithm. 
This algorithm iteratively removes those states from
which there exists an input combination for which all output combinations lead to bad states.
The resulting automaton is again represented as a symbolic automaton $A^{mpnc}$. If the initial state gets
pruned in construction, the specification is unrealizable. A counter-strategy in tree form is displayed
as explanation of unrealizability.

\item Note that in $A^{mpnc}$ each edge is labelled by a bit vector giving truth values of variables
$I \cup O \cup W$. The value of witness variable $w_i \in W$ specifies whether \qddc\/ formula $D_i^s$ holds
for all behaviours leading to this transition.

\item For each state $s$ and each input combination $ip \in 2^I$,  
we select output cum witness variable combination  $op \in 2^{O \cup W}$  such that $\delta(s,ip \cup op)$ is a valid transition of MPNC and 
$val_{ip \cup op}(\langle P_1, \ldots, P_l \rangle)$ is lexicographically maximal amongst all such $op$. 
Here, $val_{ip \cup op}(\langle P_1, \ldots, P_l \rangle)$ is $l$-bit vector with 
$i$-th bit representing truth (1 or 0) of $P_i$ under $ip \cup op$.  
If there are more than one choice of $op$ giving the same maximal value, we choose one arbitrarily. 
This gives the {\em Locally Optimal Deterministic Controller} (LODC) 
 which satisfies the lexicographically maximal subset of
soft requirements at each step. 
This greedy strategy does not guarantee global optimality.

\item The LODC can then be encoded as controller in any target language. 
We provide the encoding of LODC to LUSTRE/SCADE or NuSMV, 
which allows us to do simulation and model checking on the generated controller.
\end{itemize}

%The following example of a 2 cell arbier shows the synthesis of an arbiter module.
\begin{example}
\label{example:arbiter2cell}
Synthesis of 2 Cell Arbiter with Soft Requirements giving high priority to lower numbered request. 
Fig.~\ref{fig:2cellAutomaton1} gives the safety monitor automaton for 2-cell arbiter for following specification
\[
( \langle req_1, req_2 \rangle,  \langle ack_1, ack_2 \rangle, ARBINV ~\wedge  ArbResp(2,2), \langle \rangle, \langle ack_1, ack_2\rangle).
\]
%where $I=\{req_i\ |\ 1\leq i\leq 2\}$ and $O=\{ack_i\ |\ 1\leq i\leq 2\}$. 
%Figure \ref{fig:mtbddrepresentation} show the symbolic MTBDD representation of this monitor automaton.
Each transition is labeled by 4 bit vector giving values of $req_1, req_2, ack_1, ack_2$.
Fig.~\ref{fig:2cellAutomaton2}(a) gives the MPNC automaton for the 2-cell arbiter computed from 
the safety monitor automaton of Fig.~\ref{fig:2cellAutomaton1}. In the example, the soft requirements are $\langle ack_1, ack_2 \rangle$ 
which give $ack_1$ priority over $ack_2$. We obtain the pruned LODC controller automaton of Fig.~\ref{fig:2cellAutomaton2}(b) 
from the MPNC of Fig.~\ref{fig:2cellAutomaton2}(a). Note that we minimize the automaton at each step.
%%%%%%%%%%%%%%%%%%%%%%%%%%%%%%%%%%%%%%%%%%%%%

\begin{figure}[h]
%\begin{tabular}{p{0.6\linewidth}|p{0.25\linewidth}|p{0.15\linewidth}}
\begin{tabular}{c}
\begin{minipage}{4.8in}
\centering
\includegraphics[angle=0,scale=0.4]{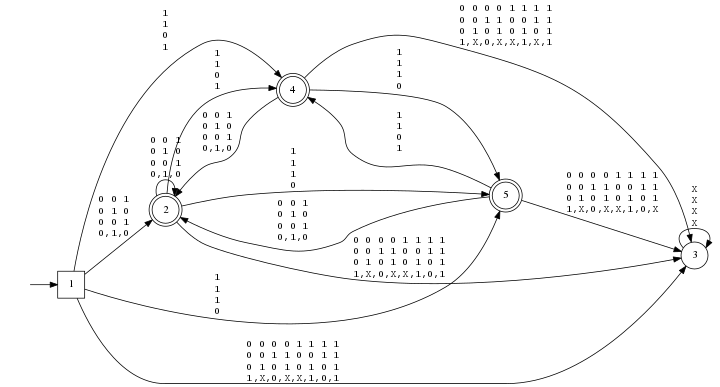}
\end{minipage}
\end{tabular}
%\vspace{-1.2\baselineskip}
\caption{\emph{Safety Monitor Automaton: 2 Cell Arbiter}}
\label{fig:2cellAutomaton1} 
\vspace{-0.25in}
\end{figure}

\begin{figure}[h]
%\begin{tabular}{p{0.6\linewidth}|p{0.25\linewidth}|p{0.15\linewidth}}
\begin{tabular}{c c}
\begin{minipage}{3in}
\centering
\includegraphics[angle=0,scale=.3]{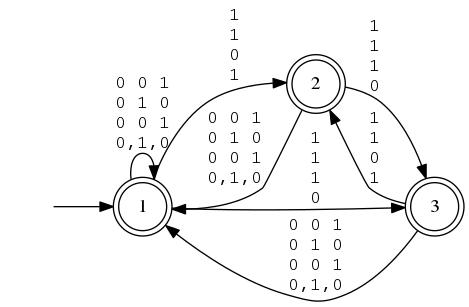}
\end{minipage}
& 
\begin{minipage}{1.8in}
\centering
\includegraphics[angle=0,scale=.3]{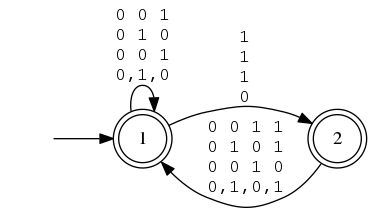}
\end{minipage}
\end{tabular}
%\vspace{-1.2\baselineskip}
\caption{(a)\emph{MPNC : 2 Cell Arbiter} (b) \emph{LODC: 2 Cell Arbiter}}
\label{fig:2cellAutomaton2} 
\vspace{-0.25in}
\end{figure}

\end{example}
%%%%%%%%%%%%%%%%%%%%%%%%%%%%%%%%%%%%%%%%%%%%%%%

%\begin{figure}[!h]
%%\begin{center}
%\centering
%\includegraphics[width=\textwidth, keepaspectratio]{fig/Arbiter2Cell_dfa.png}
%%\end{center}
%\caption{Safety Monitor Automaton: 2 Cell Arbiter}
%\label{fig:monitor2cell}
%\end{figure}

%\begin{figure}[!h]
%%\begin{center}
%\centering
%\includegraphics[width=7cm, keepaspectratio]{fig/exported_mpnc_dfa.png}
%%\end{center}
%\caption{MPNC : 2 Cell Arbiter}
%\label{fig:mpnc}
%\end{figure}

%\begin{figure}[!h]
%%\begin{center}
%\centering
%\includegraphics[width=6 cm, keepaspectratio]{fig/exported_lodc_dfa.png}
%%\end{center}
%\caption{LODC: 2 Cell Arbiter}
%\label{fig:lodc}
%\end{figure}

%amol
\paragraph{Tool Implementation} Internally, the monitor automaton, MPNC and LODC are all stored as symbolic DFA. The transition table
of the DFA is represented as MTBDD using the DFA library of tool MONA \cite{Mon01}. Internal data structures and algorithms can be found
in the full version of the paper (see Appendix \ref{subsection:algorithms}).
\vspace{-.25cm}
\subsection{Experimental Results}
 \label{subsection:experiments}
%\label{subsection:toolperformance}

Several case studies for synthesis have been carried out using \DCSYNTH. 
Table.~\ref{tab:comparisionopt} enlists the results of two of the case studies: 
% with various soft and hard requirements: 
% using \DCSYNTH\/ for all these examples: 
%The full paper gives simulations of resulting controller pointing out the impact of the soft requirements
%(see Appendix \ref{section:minepumpexample}).
\textbf{(a):} controllers for bus arbiter in \S\ref{section:motivation} 
with various soft and hard requirements, 
%(b) A {\em token ring based arbiter} specification called $Arb^{tok}(n)$ (inspired by MacMillan's arbiter) can be found in the full version of this paper. 
and \textbf{(b):} a minepump controller for the specification $MINEPUMP$ 
(cf.~Appendix. \ref{section:minepumpexample} for details). 
The controller operates a pump to get rid of water based on 
water level and methane presence (which prevents pump from being used). 
%Specification of the mine pump controller with many environmental assumptions about the period 
%and frequency of occurrenes of methane gas and assumptions about 
%the capacity of the pump to clear the water can be found in the full version of this paper  (see Appendix \ref{section:minepumpexample}). 
Soft requirements impact the quality of the pump controller. 
For example, for the soft requirement \verb#!PumpOn# the controller will 
try to keep pump {\em off} as much as possible. 
On the other hand, the soft requirement \verb#PumpOn# agressively gets rid of water by keeping 
the pump {\em on} whenever possible. 
% unless the hard requirements absolutely demand, will save power. 
%agressively gets rid of water by keeping pump on whenever possible. 
The analysis of worst case time for the controllers to get rid of water 
with different soft requirements is given in \S\ref{section:quantitative}.

\begin {table}[!h]
\caption {\DCSYNTH\/ Controller Synthesis with Soft Requirements. }
\label{tab:comparisionopt}
\begin{minipage}{\textwidth}
        \begin{center}
\resizebox{\columnwidth}{!}{%
        \begin{tabular}
        {|c|c|c|c|c|} 
                \hline
                \multirow{2}{*}{Hard Requirement}  & \multirow{2}{*}{Soft Requirement}  &  \multicolumn{3}{|c|}{ Controller Generation }\\
                \cline{3-5}
                & &  states & time (Sec) & Memory (MB) \\
                \hline      
                $ARBHARD(4,4)$ & ack4 $>>$ ... $>>$ ack1 & 50 &0.014  & 3.3 \\
                \hline    
                $ARBHARD(5,5)$ & ack5 $>>$ ... $>>$ ack1 & 432 &0.33  & 22.4 \\
                \hline  
%                $ARBHARD(5,5)$ & ack1 $>>$ ... $>>$ ack5 & 432 &0.30  & 22.4 \\
 %               \hline    
                $ARBHARD(6,6)$ &  ack6 $>>$ ... $>>$ ack1 & 4802 &14.8  & 334.5 \\
                \hline 
                $ARBINV (1 \leq i, j \leq 6)$ & Resp(req6,ack6,2)$>>$...$>>$Resp(req1,ack1,2) & 62 &  1.05 & 9.8  \\
                \hline 
                $ARBINV (1 \leq i, j \leq 5)$ & Resp(req5,ack5,3)$>>$...$>>$Resp(req1,ack1,3) & 511 &  5.4 & 32.4  \\
                \hline    
                $MINEPUMP(MPV1)$ & PumpOn $>>$ !Alarm  & 31 &0.07  & 9.1 \\
                \hline  
                $MINEPUMP(MPV2)$ &  ($CH4_{Last2Cyc}$ $=>$ !PumpOn) $>>$ PumpOn  & 34 &0.09 & 8.9 \\
                \hline
                $MINEPUMP(MPV3)$ & !PumpOn $>>$ !Alarm & 83 &0.04  & 9.1 \\
                \hline  
           
	\end{tabular}%
}
\end{center}
\end{minipage}
\end{table}

%\begin {table}[!h]
%\caption {Comparison of LODC synthesis with and without optimization}
%\label{tab:comparisionopt}
%\begin{minipage}{\textwidth}
%        \begin{center}
%        \begin{tabular}
%        {|p{0.2\linewidth} | p{0.2\linewidth} | p{0.1\linewidth}| p{0.1\linewidth}| p{0.1\linewidth}| p{0.1\linewidth}|} 
%                \hline
%                Problem &   Soft Requirements &   \multicolumn{2}{|p{0.2\linewidth}|}{ Without Optimization }    & \multicolumn{2}{|p{0.2\linewidth}|}{ With Optimization }\\
%                \hline
%                & & time (Sec) & Memory (KB)  & time (Sec) & Memory (KB) \\
%                \hline      
%                $Arb^{hard}(4,4)$ & ack4 $>>$ ... $>>$ ack1 & 0.023  & 3.4  & 0.014  & 3.3 \\
%                \hline    
%                $Arb^{hard}(5,5)$ & ack5 $>>$ ... $>>$ ack1 & 0.57 & 24.7  & 0.33  & 22.4 \\
%                \hline  
%                $Arb^{hard}(5,5)$ & ack1 $>>$ ... $>>$ ack5 & 0.54 & 25.0  & 0.30  & 22.4 \\
%                \hline    
%                $Arb^{hard}(6,6)$ &  ack6 $>>$ ... $>>$ ack1 & 24.4 & 488.8  & 14.8  & 334.5 \\
%                \hline    
%                $MP\_V1$ & PumpOn $>>$ !Alarm  & 0.0083 & 2.0  & 0.0017  & 2.1 \\
%                \hline  
%                $MP\_V2$ &  ($CH4_{Last2Cyc}$ $=>$ !PumpOn) $>>$ PumpOn & 0.0041 & 2.1  &  0.0026 & 2.2 \\
%                \hline
%                $MP\_V3$ & !PumpOn $>>$ !Alarm & 0.0053 & 2.1  & 0.0025  & 2.2 \\
%                \hline  
%           
%	\end{tabular}
%\end{center}
%\end{minipage}
%\end{table}
%
%

%paritosh
\vspace{-1cm}
\section{Robustness}
\label{section:robustness}

We consider various notions of robustness. 
Table.~\ref{tab:robustformula} summarizes the robustness that we consider 
along with hard and soft requirements to obtain robust controllers.
\begin{itemize}
 \item {\bf Be-Correct} If assumption has held invariantly so far then commitment must hold now.
 \item {\bf Be-Currently-Correct} If assumption holds intermittently, the commitment must hold whenever assumption holds.
 \item {\bf Degraded-Performance} Let \verb#Ad# $\subseteq$ \verb#A# and  \verb#Cd# $\subseteq$ \verb#C#, where
 \verb#Ad,Cd# denote reduced set of assumptions and commitments which specify degraded behaviour of system when fewer assumptions
 hold.
 \item {\bf Never-Give-Up} In addition to Be-Correct, all the commitments are asserted as soft requirements. This makes the controller
 synthesizer attempt to make them true even when assumptions do not hold, and for as many inputs as possible. 
 \item {\bf Greedy} Here commitments are given as soft goals ignoring the assumptions. The synthesis algorithm tries ot make as many commitments
 true as possible at each step in a greedy fashion.
\end{itemize}
%\medskip

Resilient synthesis requires synthesis of controller which works under weaker assumptions than Be-Correct notion in order to tolerate errors 
in assumptions. For example, Be-Currently-Correct requires commitment to hold at now if assumption holds now irrespective of whether it has held in past.
Several other notions of resilience are given below. A notion of $k,b$-resilience was proposed by Ehler and Topcu \cite{ET14}, 
and further adapted by Bloem as
$k$-robustness \cite{BKKW15}.
\begin{table}
\caption{Robust synthesis notions and their specifications.}
\label{tab:robustformula}
In the table \verb#A# denotes conjunction of assumptions
and \verb#C# denotes conjunction of commitments. Thus, \verb#!A# denotes violation of at least one assumption.
\begin{tabular}{|l|c|c|}
\hline
Robustness Criterion  &  Hard Requirement  & Soft Requirement 
\\ \hline\hline
Be-Correct(A,C)  			&  \verb#G(Pref(A) => C)#	&   
\\ \hline
Be-Currently-Correct(A,C)		& \verb#G(A => C)#  &
\\ \hline
Degraded-Performance(A,C,Ad,Cd)		& \verb#G((A => C) && (Ad => Cd)# & 
\\ \hline
Never-Give-Up(A,C)		& \verb#G(A => C)#	&  \verb#C# 
\\ \hline
Greedy(C)				&			&  \verb#C# 
\\ \hline
$k$-Bounded(A,C,k)			& \verb#G((scount !A < k) => C)#  &
\\ \hline 
$k,b$-Resilient(A,C,k,b)			& \verb#G(KBREZ(A) => C)#  &
\\ \hline 
$k,b$-Variant(A,C,k,b)			& \verb#G([]((slen = b &&#  &
\\  
&\verb#scount !A <= k) => C))#  &
\\ \hline 
%$k,b$-variant2			& \verb#G((!(true^(slen = b & (scount !A > k)))) => C)#  &
%\\ \hline 
\end{tabular}

where \verb#KBREZ(A) = !(true^((scount !A >= k) && []([A] => slen < b))^true)#
\end{table}

\vspace{-0.5cm}
\begin{itemize}
 \item {\bf $k$-Bounded} If in past assumptions have been violated at most $k$ times so far then commitment must be met. 
 \item {\bf $k,b$-Resilient} A subinterval where assumption is continuously true for $b$ or more cycles is called a recovery period. 
 Formula \verb#KBREZ(A)# in Table.~\ref{tab:robustformula} states that between any two recovery
 periods, the maximum number of assumption violations is at most $k$. A controller which guarantees commitments $C$ at every point where past satisfies
 \verb#KBREZ(A)# is called $k,b$-resilient (see \cite{ET14}).
 \item {\bf $k,b$-Variant} If in past in any period of length $b$ the assumption has been violated at most $k$ times then the commitment must hold.
% \item {\bf $k,b$-variant2} Unless we see violation of assumption more than $k$ times in last $b$ cycles, the commitment must be met. 
% Past before last $b$ cycles does not matter. 
\end{itemize}
Note that criterion such as Never-Give-Up can be combined with Resilient synthesis. Moreover, designer may selectively apply these criteria to 
specific assumptions and commitments. \DCSYNTH\/ permits full flexibility in making such choices.
Table.~\ref{tab:robustArbiterSynthesis} gives the synthesis of 
arbiter specifications %in \S.~\ref{section:arbiterexample}
 under various notions of robustness in \DCSYNTH\/ for the assumptions $Assume(n,i)$ 
and commitments $ARBHARD(n,k)$ for the specification $Arb^{hard}_{assume}(n,k,i)$ in Equation.~\ref{equation5}. 
\begin{table}
\caption{Synthesis of Robust Arbiters in DCSynth}
\label{tab:robustArbiterSynthesis}
%\centering
\resizebox{\columnwidth}{!}{%
\begin{tabular}{|l|c|c|c|c|}
\hline
 \multicolumn{2}{|c|}{Specification} & \multicolumn{2}{c|}{LODC} \\
 \cline{1-2}
 \cline{3-4}
 Robustness Specification & Priorities of Soft Requirements & States & Time\\
 \hline
 \hline
 Be-Correct($Assume(4,2), ARBHARD(4,2)$) & - &6 & 0.02\\
 \hline
 Be-Currently-Correct($Assume(4,2), ARBHARD(4,2)$) & - &Unrealizable &\\
 \hline
 Never-Give-Up($Assume(4,2), ARBHARD(4,2)$) &$(ARBINV >>$ $Resp(req_1,ack_1,2)$ &15 &0.73\\
 & $ >>\ldots >>Resp(req_4,ack_4,2)$ &&\\
 \hline
 Greedy($ARBHARD(4,2)$) & $(ARBINV >>$ $Resp(req_1,ack_1,2)$  & 15 & 0.07 \\
 & $ \ldots $ $>>Resp(req_4,ack_4,2)$ & & \\
 \hline
 $k$-Bounded($Assume(4,2), ARBHARD(4,3),2$) & - &29 & 0.06 \\
 \hline
 $k,b$-Resilient($Assume(4,2), ARBHARD(4,3),2,3)$ &- &39  &0.09  \\
 \hline
  $k,b$-Variant($Assume(4,2), ARBHARD(4,3),2,3$) &-   &27 & 0.22 \\
 \hline
% $k,b$-variant2 & Unrealizable & - &  &  \\
% \hline
\end{tabular}%
}
\end{table}

\begin{figure}
% Sim2chro version 3.4 beta (c) Sep 2000
% please report bugs, so I can report them as features ;-)
{
\def\lignefine{\linethickness{0.05pt}}
\def\ligneepaisse{\linethickness{2pt}}
\noindent
\setlength{\unitlength}{1mm}
\begin{picture}(136.0000,60.0000)(-5.0000,0.0000)
\fboxsep 0pt
\lignefine
\color{black}
\multiput(0.0000,-5.0000)(4.0000,0){30}{\line(0,1){60.0000}}
\put(2.0000,54.0000){\scriptsize\makebox(0,0)[t]{1}}
\put(2.0000,-4.0000){\scriptsize\makebox(0,0)[b]{1}}
\put(6.0000,54.0000){\scriptsize\makebox(0,0)[t]{2}}
\put(6.0000,-4.0000){\scriptsize\makebox(0,0)[b]{2}}
\put(10.0000,54.0000){\scriptsize\makebox(0,0)[t]{3}}
\put(10.0000,-4.0000){\scriptsize\makebox(0,0)[b]{3}}
\put(14.0000,54.0000){\scriptsize\makebox(0,0)[t]{4}}
\put(14.0000,-4.0000){\scriptsize\makebox(0,0)[b]{4}}
\put(18.0000,54.0000){\scriptsize\makebox(0,0)[t]{5}}
\put(18.0000,-4.0000){\scriptsize\makebox(0,0)[b]{5}}
\put(22.0000,54.0000){\scriptsize\makebox(0,0)[t]{6}}
\put(22.0000,-4.0000){\scriptsize\makebox(0,0)[b]{6}}
\put(26.0000,54.0000){\scriptsize\makebox(0,0)[t]{7}}
\put(26.0000,-4.0000){\scriptsize\makebox(0,0)[b]{7}}
\put(30.0000,54.0000){\scriptsize\makebox(0,0)[t]{8}}
\put(30.0000,-4.0000){\scriptsize\makebox(0,0)[b]{8}}
\put(34.0000,54.0000){\scriptsize\makebox(0,0)[t]{9}}
\put(34.0000,-4.0000){\scriptsize\makebox(0,0)[b]{9}}
\put(38.0000,54.0000){\scriptsize\makebox(0,0)[t]{10}}
\put(38.0000,-4.0000){\scriptsize\makebox(0,0)[b]{10}}
\put(42.0000,54.0000){\scriptsize\makebox(0,0)[t]{11}}
\put(42.0000,-4.0000){\scriptsize\makebox(0,0)[b]{11}}
\put(46.0000,54.0000){\scriptsize\makebox(0,0)[t]{12}}
\put(46.0000,-4.0000){\scriptsize\makebox(0,0)[b]{12}}
\put(50.0000,54.0000){\scriptsize\makebox(0,0)[t]{13}}
\put(50.0000,-4.0000){\scriptsize\makebox(0,0)[b]{13}}
\put(54.0000,54.0000){\scriptsize\makebox(0,0)[t]{14}}
\put(54.0000,-4.0000){\scriptsize\makebox(0,0)[b]{14}}
\put(58.0000,54.0000){\scriptsize\makebox(0,0)[t]{15}}
\put(58.0000,-4.0000){\scriptsize\makebox(0,0)[b]{15}}
\put(62.0000,54.0000){\scriptsize\makebox(0,0)[t]{16}}
\put(62.0000,-4.0000){\scriptsize\makebox(0,0)[b]{16}}
\put(66.0000,54.0000){\scriptsize\makebox(0,0)[t]{17}}
\put(66.0000,-4.0000){\scriptsize\makebox(0,0)[b]{17}}
\put(70.0000,54.0000){\scriptsize\makebox(0,0)[t]{18}}
\put(70.0000,-4.0000){\scriptsize\makebox(0,0)[b]{18}}
\put(74.0000,54.0000){\scriptsize\makebox(0,0)[t]{19}}
\put(74.0000,-4.0000){\scriptsize\makebox(0,0)[b]{19}}
\put(78.0000,54.0000){\scriptsize\makebox(0,0)[t]{20}}
\put(78.0000,-4.0000){\scriptsize\makebox(0,0)[b]{20}}
\put(82.0000,54.0000){\scriptsize\makebox(0,0)[t]{21}}
\put(82.0000,-4.0000){\scriptsize\makebox(0,0)[b]{21}}
\put(86.0000,54.0000){\scriptsize\makebox(0,0)[t]{22}}
\put(86.0000,-4.0000){\scriptsize\makebox(0,0)[b]{22}}
\put(90.0000,54.0000){\scriptsize\makebox(0,0)[t]{23}}
\put(90.0000,-4.0000){\scriptsize\makebox(0,0)[b]{23}}
\put(94.0000,54.0000){\scriptsize\makebox(0,0)[t]{24}}
\put(94.0000,-4.0000){\scriptsize\makebox(0,0)[b]{24}}
\put(98.0000,54.0000){\scriptsize\makebox(0,0)[t]{25}}
\put(98.0000,-4.0000){\scriptsize\makebox(0,0)[b]{25}}
\put(102.0000,54.0000){\scriptsize\makebox(0,0)[t]{26}}
\put(102.0000,-4.0000){\scriptsize\makebox(0,0)[b]{26}}
\put(106.0000,54.0000){\scriptsize\makebox(0,0)[t]{27}}
\put(106.0000,-4.0000){\scriptsize\makebox(0,0)[b]{27}}
\put(110.0000,54.0000){\scriptsize\makebox(0,0)[t]{28}}
\put(110.0000,-4.0000){\scriptsize\makebox(0,0)[b]{28}}
\put(114.0000,54.0000){\scriptsize\makebox(0,0)[t]{29}}
\put(114.0000,-4.0000){\scriptsize\makebox(0,0)[b]{29}}
\put(-1.0000,44.0000){\line(1,0){118.0000}}
\put(-1.0000,48.0000){\line(1,0){118.0000}}
\put(-1.0000,38.0000){\line(1,0){118.0000}}
\put(-1.0000,42.0000){\line(1,0){118.0000}}
\put(-1.0000,32.0000){\line(1,0){118.0000}}
\put(-1.0000,36.0000){\line(1,0){118.0000}}
\put(-1.0000,26.0000){\line(1,0){118.0000}}
\put(-1.0000,30.0000){\line(1,0){118.0000}}
\put(-1.0000,20.0000){\line(1,0){118.0000}}
\put(-1.0000,24.0000){\line(1,0){118.0000}}
\put(-1.0000,14.0000){\line(1,0){118.0000}}
\put(-1.0000,18.0000){\line(1,0){118.0000}}
\put(-1.0000,8.0000){\line(1,0){118.0000}}
\put(-1.0000,12.0000){\line(1,0){118.0000}}
\put(-1.0000,2.0000){\line(1,0){118.0000}}
\put(-1.0000,6.0000){\line(1,0){118.0000}}
\ligneepaisse
\color{blue}
\put(-1.0000,46.0000){\color{blue}\normalsize\makebox(0,0)[r]{req1}}
\put(12.0000,44.0000){\line(0,1){4.0000}}
\put(16.0000,48.0000){\line(0,-1){4.0000}}
\put(36.0000,44.0000){\line(0,1){4.0000}}
\put(48.0000,48.0000){\line(0,-1){4.0000}}
\put(72.0000,44.0000){\line(0,1){4.0000}}
\put(100.0000,48.0000){\line(0,-1){4.0000}}
\put(0.0000,44.0000){\line(1,0){4.0000}}
\put(4.0000,44.0000){\line(1,0){4.0000}}
\put(8.0000,44.0000){\line(1,0){4.0000}}
\put(12.0000,48.0000){\line(1,0){4.0000}}
\put(16.0000,44.0000){\line(1,0){4.0000}}
\put(20.0000,44.0000){\line(1,0){4.0000}}
\put(24.0000,44.0000){\line(1,0){4.0000}}
\put(28.0000,44.0000){\line(1,0){4.0000}}
\put(32.0000,44.0000){\line(1,0){4.0000}}
\put(36.0000,48.0000){\line(1,0){4.0000}}
\put(40.0000,48.0000){\line(1,0){4.0000}}
\put(44.0000,48.0000){\line(1,0){4.0000}}
\put(48.0000,44.0000){\line(1,0){4.0000}}
\put(52.0000,44.0000){\line(1,0){4.0000}}
\put(56.0000,44.0000){\line(1,0){4.0000}}
\put(60.0000,44.0000){\line(1,0){4.0000}}
\put(64.0000,44.0000){\line(1,0){4.0000}}
\put(68.0000,44.0000){\line(1,0){4.0000}}
\put(72.0000,48.0000){\line(1,0){4.0000}}
\put(76.0000,48.0000){\line(1,0){4.0000}}
\put(80.0000,48.0000){\line(1,0){4.0000}}
\put(84.0000,48.0000){\line(1,0){4.0000}}
\put(88.0000,48.0000){\line(1,0){4.0000}}
\put(92.0000,48.0000){\line(1,0){4.0000}}
\put(96.0000,48.0000){\line(1,0){4.0000}}
\put(100.0000,44.0000){\line(1,0){4.0000}}
\put(104.0000,44.0000){\line(1,0){4.0000}}
\put(108.0000,44.0000){\line(1,0){4.0000}}
\put(112.0000,44.0000){\line(1,0){4.0000}}
\color{blue}
\put(-1.0000,40.0000){\color{blue}\normalsize\makebox(0,0)[r]{req2}}
\put(4.0000,38.0000){\line(0,1){4.0000}}
\put(8.0000,42.0000){\line(0,-1){4.0000}}
\put(20.0000,38.0000){\line(0,1){4.0000}}
\put(36.0000,42.0000){\line(0,-1){4.0000}}
\put(48.0000,38.0000){\line(0,1){4.0000}}
\put(0.0000,38.0000){\line(1,0){4.0000}}
\put(4.0000,42.0000){\line(1,0){4.0000}}
\put(8.0000,38.0000){\line(1,0){4.0000}}
\put(12.0000,38.0000){\line(1,0){4.0000}}
\put(16.0000,38.0000){\line(1,0){4.0000}}
\put(20.0000,42.0000){\line(1,0){4.0000}}
\put(24.0000,42.0000){\line(1,0){4.0000}}
\put(28.0000,42.0000){\line(1,0){4.0000}}
\put(32.0000,42.0000){\line(1,0){4.0000}}
\put(36.0000,38.0000){\line(1,0){4.0000}}
\put(40.0000,38.0000){\line(1,0){4.0000}}
\put(44.0000,38.0000){\line(1,0){4.0000}}
\put(48.0000,42.0000){\line(1,0){4.0000}}
\put(52.0000,42.0000){\line(1,0){4.0000}}
\put(56.0000,42.0000){\line(1,0){4.0000}}
\put(60.0000,42.0000){\line(1,0){4.0000}}
\put(64.0000,42.0000){\line(1,0){4.0000}}
\put(68.0000,42.0000){\line(1,0){4.0000}}
\put(72.0000,42.0000){\line(1,0){4.0000}}
\put(76.0000,42.0000){\line(1,0){4.0000}}
\put(80.0000,42.0000){\line(1,0){4.0000}}
\put(84.0000,42.0000){\line(1,0){4.0000}}
\put(88.0000,42.0000){\line(1,0){4.0000}}
\put(92.0000,42.0000){\line(1,0){4.0000}}
\put(96.0000,42.0000){\line(1,0){4.0000}}
\put(100.0000,42.0000){\line(1,0){4.0000}}
\put(104.0000,42.0000){\line(1,0){4.0000}}
\put(108.0000,42.0000){\line(1,0){4.0000}}
\put(112.0000,42.0000){\line(1,0){4.0000}}
\color{blue}
\put(-1.0000,34.0000){\color{blue}\normalsize\makebox(0,0)[r]{req3}}
\put(8.0000,32.0000){\line(0,1){4.0000}}
\put(12.0000,36.0000){\line(0,-1){4.0000}}
\put(48.0000,32.0000){\line(0,1){4.0000}}
\put(88.0000,36.0000){\line(0,-1){4.0000}}
\put(0.0000,32.0000){\line(1,0){4.0000}}
\put(4.0000,32.0000){\line(1,0){4.0000}}
\put(8.0000,36.0000){\line(1,0){4.0000}}
\put(12.0000,32.0000){\line(1,0){4.0000}}
\put(16.0000,32.0000){\line(1,0){4.0000}}
\put(20.0000,32.0000){\line(1,0){4.0000}}
\put(24.0000,32.0000){\line(1,0){4.0000}}
\put(28.0000,32.0000){\line(1,0){4.0000}}
\put(32.0000,32.0000){\line(1,0){4.0000}}
\put(36.0000,32.0000){\line(1,0){4.0000}}
\put(40.0000,32.0000){\line(1,0){4.0000}}
\put(44.0000,32.0000){\line(1,0){4.0000}}
\put(48.0000,36.0000){\line(1,0){4.0000}}
\put(52.0000,36.0000){\line(1,0){4.0000}}
\put(56.0000,36.0000){\line(1,0){4.0000}}
\put(60.0000,36.0000){\line(1,0){4.0000}}
\put(64.0000,36.0000){\line(1,0){4.0000}}
\put(68.0000,36.0000){\line(1,0){4.0000}}
\put(72.0000,36.0000){\line(1,0){4.0000}}
\put(76.0000,36.0000){\line(1,0){4.0000}}
\put(80.0000,36.0000){\line(1,0){4.0000}}
\put(84.0000,36.0000){\line(1,0){4.0000}}
\put(88.0000,32.0000){\line(1,0){4.0000}}
\put(92.0000,32.0000){\line(1,0){4.0000}}
\put(96.0000,32.0000){\line(1,0){4.0000}}
\put(100.0000,32.0000){\line(1,0){4.0000}}
\put(104.0000,32.0000){\line(1,0){4.0000}}
\put(108.0000,32.0000){\line(1,0){4.0000}}
\put(112.0000,32.0000){\line(1,0){4.0000}}
\color{blue}
\put(-1.0000,28.0000){\color{blue}\normalsize\makebox(0,0)[r]{req4}}
\put(16.0000,26.0000){\line(0,1){4.0000}}
\put(48.0000,30.0000){\line(0,-1){4.0000}}
\put(56.0000,26.0000){\line(0,1){4.0000}}
\put(112.0000,30.0000){\line(0,-1){4.0000}}
\put(0.0000,26.0000){\line(1,0){4.0000}}
\put(4.0000,26.0000){\line(1,0){4.0000}}
\put(8.0000,26.0000){\line(1,0){4.0000}}
\put(12.0000,26.0000){\line(1,0){4.0000}}
\put(16.0000,30.0000){\line(1,0){4.0000}}
\put(20.0000,30.0000){\line(1,0){4.0000}}
\put(24.0000,30.0000){\line(1,0){4.0000}}
\put(28.0000,30.0000){\line(1,0){4.0000}}
\put(32.0000,30.0000){\line(1,0){4.0000}}
\put(36.0000,30.0000){\line(1,0){4.0000}}
\put(40.0000,30.0000){\line(1,0){4.0000}}
\put(44.0000,30.0000){\line(1,0){4.0000}}
\put(48.0000,26.0000){\line(1,0){4.0000}}
\put(52.0000,26.0000){\line(1,0){4.0000}}
\put(56.0000,30.0000){\line(1,0){4.0000}}
\put(60.0000,30.0000){\line(1,0){4.0000}}
\put(64.0000,30.0000){\line(1,0){4.0000}}
\put(68.0000,30.0000){\line(1,0){4.0000}}
\put(72.0000,30.0000){\line(1,0){4.0000}}
\put(76.0000,30.0000){\line(1,0){4.0000}}
\put(80.0000,30.0000){\line(1,0){4.0000}}
\put(84.0000,30.0000){\line(1,0){4.0000}}
\put(88.0000,30.0000){\line(1,0){4.0000}}
\put(92.0000,30.0000){\line(1,0){4.0000}}
\put(96.0000,30.0000){\line(1,0){4.0000}}
\put(100.0000,30.0000){\line(1,0){4.0000}}
\put(104.0000,30.0000){\line(1,0){4.0000}}
\put(108.0000,30.0000){\line(1,0){4.0000}}
\put(112.0000,26.0000){\line(1,0){4.0000}}
\color{red}
\put(-1.0000,22.0000){\color{red}\normalsize\makebox(0,0)[r]{ack1}}
\put(12.0000,20.0000){\line(0,1){4.0000}}
\put(16.0000,24.0000){\line(0,-1){4.0000}}
\put(36.0000,20.0000){\line(0,1){4.0000}}
\put(40.0000,24.0000){\line(0,-1){4.0000}}
\put(44.0000,20.0000){\line(0,1){4.0000}}
\put(48.0000,24.0000){\line(0,-1){4.0000}}
\put(76.0000,20.0000){\line(0,1){4.0000}}
\put(80.0000,24.0000){\line(0,-1){4.0000}}
\put(84.0000,20.0000){\line(0,1){4.0000}}
\put(88.0000,24.0000){\line(0,-1){4.0000}}
\put(92.0000,20.0000){\line(0,1){4.0000}}
\put(96.0000,24.0000){\line(0,-1){4.0000}}
\put(0.0000,20.0000){\line(1,0){4.0000}}
\put(4.0000,20.0000){\line(1,0){4.0000}}
\put(8.0000,20.0000){\line(1,0){4.0000}}
\put(12.0000,24.0000){\line(1,0){4.0000}}
\put(16.0000,20.0000){\line(1,0){4.0000}}
\put(20.0000,20.0000){\line(1,0){4.0000}}
\put(24.0000,20.0000){\line(1,0){4.0000}}
\put(28.0000,20.0000){\line(1,0){4.0000}}
\put(32.0000,20.0000){\line(1,0){4.0000}}
\put(36.0000,24.0000){\line(1,0){4.0000}}
\put(40.0000,20.0000){\line(1,0){4.0000}}
\put(44.0000,24.0000){\line(1,0){4.0000}}
\put(48.0000,20.0000){\line(1,0){4.0000}}
\put(52.0000,20.0000){\line(1,0){4.0000}}
\put(56.0000,20.0000){\line(1,0){4.0000}}
\put(60.0000,20.0000){\line(1,0){4.0000}}
\put(64.0000,20.0000){\line(1,0){4.0000}}
\put(68.0000,20.0000){\line(1,0){4.0000}}
\put(72.0000,20.0000){\line(1,0){4.0000}}
\put(76.0000,24.0000){\line(1,0){4.0000}}
\put(80.0000,20.0000){\line(1,0){4.0000}}
\put(84.0000,24.0000){\line(1,0){4.0000}}
\put(88.0000,20.0000){\line(1,0){4.0000}}
\put(92.0000,24.0000){\line(1,0){4.0000}}
\put(96.0000,20.0000){\line(1,0){4.0000}}
\put(100.0000,20.0000){\line(1,0){4.0000}}
\put(104.0000,20.0000){\line(1,0){4.0000}}
\put(108.0000,20.0000){\line(1,0){4.0000}}
\put(112.0000,20.0000){\line(1,0){4.0000}}
\color{red}
\put(-1.0000,16.0000){\color{red}\normalsize\makebox(0,0)[r]{ack2}}
\put(4.0000,14.0000){\line(0,1){4.0000}}
\put(8.0000,18.0000){\line(0,-1){4.0000}}
\put(20.0000,14.0000){\line(0,1){4.0000}}
\put(24.0000,18.0000){\line(0,-1){4.0000}}
\put(28.0000,14.0000){\line(0,1){4.0000}}
\put(32.0000,18.0000){\line(0,-1){4.0000}}
\put(48.0000,14.0000){\line(0,1){4.0000}}
\put(52.0000,18.0000){\line(0,-1){4.0000}}
\put(56.0000,14.0000){\line(0,1){4.0000}}
\put(60.0000,18.0000){\line(0,-1){4.0000}}
\put(64.0000,14.0000){\line(0,1){4.0000}}
\put(68.0000,18.0000){\line(0,-1){4.0000}}
\put(72.0000,14.0000){\line(0,1){4.0000}}
\put(76.0000,18.0000){\line(0,-1){4.0000}}
\put(80.0000,14.0000){\line(0,1){4.0000}}
\put(84.0000,18.0000){\line(0,-1){4.0000}}
\put(88.0000,14.0000){\line(0,1){4.0000}}
\put(92.0000,18.0000){\line(0,-1){4.0000}}
\put(96.0000,14.0000){\line(0,1){4.0000}}
\put(100.0000,18.0000){\line(0,-1){4.0000}}
\put(104.0000,14.0000){\line(0,1){4.0000}}
\put(108.0000,18.0000){\line(0,-1){4.0000}}
\put(112.0000,14.0000){\line(0,1){4.0000}}
\put(0.0000,14.0000){\line(1,0){4.0000}}
\put(4.0000,18.0000){\line(1,0){4.0000}}
\put(8.0000,14.0000){\line(1,0){4.0000}}
\put(12.0000,14.0000){\line(1,0){4.0000}}
\put(16.0000,14.0000){\line(1,0){4.0000}}
\put(20.0000,18.0000){\line(1,0){4.0000}}
\put(24.0000,14.0000){\line(1,0){4.0000}}
\put(28.0000,18.0000){\line(1,0){4.0000}}
\put(32.0000,14.0000){\line(1,0){4.0000}}
\put(36.0000,14.0000){\line(1,0){4.0000}}
\put(40.0000,14.0000){\line(1,0){4.0000}}
\put(44.0000,14.0000){\line(1,0){4.0000}}
\put(48.0000,18.0000){\line(1,0){4.0000}}
\put(52.0000,14.0000){\line(1,0){4.0000}}
\put(56.0000,18.0000){\line(1,0){4.0000}}
\put(60.0000,14.0000){\line(1,0){4.0000}}
\put(64.0000,18.0000){\line(1,0){4.0000}}
\put(68.0000,14.0000){\line(1,0){4.0000}}
\put(72.0000,18.0000){\line(1,0){4.0000}}
\put(76.0000,14.0000){\line(1,0){4.0000}}
\put(80.0000,18.0000){\line(1,0){4.0000}}
\put(84.0000,14.0000){\line(1,0){4.0000}}
\put(88.0000,18.0000){\line(1,0){4.0000}}
\put(92.0000,14.0000){\line(1,0){4.0000}}
\put(96.0000,18.0000){\line(1,0){4.0000}}
\put(100.0000,14.0000){\line(1,0){4.0000}}
\put(104.0000,18.0000){\line(1,0){4.0000}}
\put(108.0000,14.0000){\line(1,0){4.0000}}
\put(112.0000,18.0000){\line(1,0){4.0000}}
\color{red}
\put(-1.0000,10.0000){\color{red}\normalsize\makebox(0,0)[r]{ack3}}
\put(8.0000,8.0000){\line(0,1){4.0000}}
\put(12.0000,12.0000){\line(0,-1){4.0000}}
\put(52.0000,8.0000){\line(0,1){4.0000}}
\put(56.0000,12.0000){\line(0,-1){4.0000}}
\put(60.0000,8.0000){\line(0,1){4.0000}}
\put(64.0000,12.0000){\line(0,-1){4.0000}}
\put(68.0000,8.0000){\line(0,1){4.0000}}
\put(72.0000,12.0000){\line(0,-1){4.0000}}
\put(0.0000,8.0000){\line(1,0){4.0000}}
\put(4.0000,8.0000){\line(1,0){4.0000}}
\put(8.0000,12.0000){\line(1,0){4.0000}}
\put(12.0000,8.0000){\line(1,0){4.0000}}
\put(16.0000,8.0000){\line(1,0){4.0000}}
\put(20.0000,8.0000){\line(1,0){4.0000}}
\put(24.0000,8.0000){\line(1,0){4.0000}}
\put(28.0000,8.0000){\line(1,0){4.0000}}
\put(32.0000,8.0000){\line(1,0){4.0000}}
\put(36.0000,8.0000){\line(1,0){4.0000}}
\put(40.0000,8.0000){\line(1,0){4.0000}}
\put(44.0000,8.0000){\line(1,0){4.0000}}
\put(48.0000,8.0000){\line(1,0){4.0000}}
\put(52.0000,12.0000){\line(1,0){4.0000}}
\put(56.0000,8.0000){\line(1,0){4.0000}}
\put(60.0000,12.0000){\line(1,0){4.0000}}
\put(64.0000,8.0000){\line(1,0){4.0000}}
\put(68.0000,12.0000){\line(1,0){4.0000}}
\put(72.0000,8.0000){\line(1,0){4.0000}}
\put(76.0000,8.0000){\line(1,0){4.0000}}
\put(80.0000,8.0000){\line(1,0){4.0000}}
\put(84.0000,8.0000){\line(1,0){4.0000}}
\put(88.0000,8.0000){\line(1,0){4.0000}}
\put(92.0000,8.0000){\line(1,0){4.0000}}
\put(96.0000,8.0000){\line(1,0){4.0000}}
\put(100.0000,8.0000){\line(1,0){4.0000}}
\put(104.0000,8.0000){\line(1,0){4.0000}}
\put(108.0000,8.0000){\line(1,0){4.0000}}
\put(112.0000,8.0000){\line(1,0){4.0000}}
\color{red}
\put(-1.0000,4.0000){\color{red}\normalsize\makebox(0,0)[r]{ack4}}
\put(16.0000,2.0000){\line(0,1){4.0000}}
\put(20.0000,6.0000){\line(0,-1){4.0000}}
\put(24.0000,2.0000){\line(0,1){4.0000}}
\put(28.0000,6.0000){\line(0,-1){4.0000}}
\put(32.0000,2.0000){\line(0,1){4.0000}}
\put(36.0000,6.0000){\line(0,-1){4.0000}}
\put(40.0000,2.0000){\line(0,1){4.0000}}
\put(44.0000,6.0000){\line(0,-1){4.0000}}
\put(100.0000,2.0000){\line(0,1){4.0000}}
\put(104.0000,6.0000){\line(0,-1){4.0000}}
\put(108.0000,2.0000){\line(0,1){4.0000}}
\put(112.0000,6.0000){\line(0,-1){4.0000}}
\put(0.0000,2.0000){\line(1,0){4.0000}}
\put(4.0000,2.0000){\line(1,0){4.0000}}
\put(8.0000,2.0000){\line(1,0){4.0000}}
\put(12.0000,2.0000){\line(1,0){4.0000}}
\put(16.0000,6.0000){\line(1,0){4.0000}}
\put(20.0000,2.0000){\line(1,0){4.0000}}
\put(24.0000,6.0000){\line(1,0){4.0000}}
\put(28.0000,2.0000){\line(1,0){4.0000}}
\put(32.0000,6.0000){\line(1,0){4.0000}}
\put(36.0000,2.0000){\line(1,0){4.0000}}
\put(40.0000,6.0000){\line(1,0){4.0000}}
\put(44.0000,2.0000){\line(1,0){4.0000}}
\put(48.0000,2.0000){\line(1,0){4.0000}}
\put(52.0000,2.0000){\line(1,0){4.0000}}
\put(56.0000,2.0000){\line(1,0){4.0000}}
\put(60.0000,2.0000){\line(1,0){4.0000}}
\put(64.0000,2.0000){\line(1,0){4.0000}}
\put(68.0000,2.0000){\line(1,0){4.0000}}
\put(72.0000,2.0000){\line(1,0){4.0000}}
\put(76.0000,2.0000){\line(1,0){4.0000}}
\put(80.0000,2.0000){\line(1,0){4.0000}}
\put(84.0000,2.0000){\line(1,0){4.0000}}
\put(88.0000,2.0000){\line(1,0){4.0000}}
\put(92.0000,2.0000){\line(1,0){4.0000}}
\put(96.0000,2.0000){\line(1,0){4.0000}}
\put(100.0000,6.0000){\line(1,0){4.0000}}
\put(104.0000,2.0000){\line(1,0){4.0000}}
\put(108.0000,6.0000){\line(1,0){4.0000}}
\put(112.0000,2.0000){\line(1,0){4.0000}}
%\caption{some caption} 
\end{picture}
}
\caption{Example Simulation of Robust Controller for Never-Give-Up}
\label{fig:neverGiveUpSim}
\end{figure}
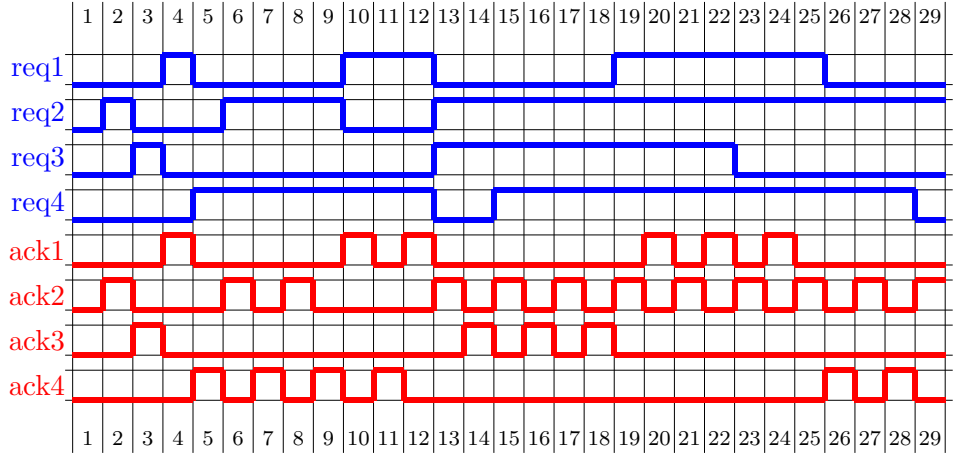

The simulation of controller produced for Never-Give-Up Strategy is given is Figure
\ref{fig:neverGiveUpSim}. The assumtions starts violating after step 15,
where more than request are true simultaneously, but the controller tries to meet
as many requirements as possible.

%amol
%\vspace{-1cm}
\section{Shield Synthesis}
\label{section:shieldsynthesis}
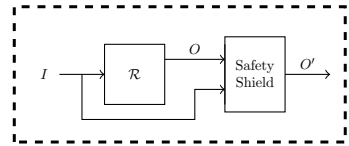
\begin{wrapfigure}{r}{.4\textwidth}
%\begin{figure}[!h]
\begin{center}
\begin{tikzpicture}[scale=0.2, every node/.style={scale=0.5}]
\draw (2,0) node {$I$};
\draw[->] (3,0) -- (6,0);
\draw (6,-2) rectangle (10,2);
\draw (8,0) node {$\R$};
\draw[->] (10,1) -- (14,1) node[midway,above] {$O$};
\draw[->] (4.5,0) -- (4.5,-3) -- (12,-3) -- (12,-1) -- (14,-1);
\draw (14,-2) -- (14, 2);
\draw (14,-2.5) rectangle (18,2.5);
\draw (16,0.5) node {Safety} (16,-0.5) node {Shield};
\draw[->] (18,0) -- (21,0) node[midway, above] {$O'$};
%\draw (2,3.5) node {$D$};
%\draw[->] (3,3.5) -- (16,3.5) -- (16, 2.5);
\draw[line width=.4mm, dashed] (0,-4.5) rectangle (22,4.5); 
\end{tikzpicture}
\end{center}
\caption{Safety shield.}
\label{fig:safety-shield}
%\end{figure}
\end{wrapfigure}
%Reactive synthesis against a temporal specification is often hard in practice 
%due to high time and space complexity. 
%due to memory requirement and performance reasons. 
%However, in many cases one could synthesize a {\em shield} 
%for a small number of (critical) properties 
%which when attached to a reactive system can detect the property violations by the system 
%and correct them {\em in situ} or {\em in due course of time}. 
A {\em safety shield} is a run time enforcer that can be attached to 
a reactive system design $\R$ to detect the property violations by the design 
and correct them run time \cite{BKKW15, WWZ16}. 
Fig.~\ref{fig:safety-shield} gives the schematic of a safety shield for $\R$.

We assume the definitions of a reactive system design and their serial composition 
(cf.~\cite{WWZ16} for details). 
Let $D$ be a \qddc formula over $I\union O$. 
%We say $\R$ {\em satisfies} the formula $D$, written $\R\models D$, if $L(\R)\subseteq L(D)$. 
Let $O=\{o_1, \cdots, o_n\}\subseteq\Sigma$ and let $O'=\{o_1',\ldots, o_n'\}$. 
Then we define $D[O/O']$ to be the formula $D'$ obtained by replacing 
every occurrence of $o_i$ in $D$ by $o_i'$ for all $1\leq i\leq n$. 
We can now define a safety shield. 
\begin{definition}[Safety shield]\cite{BKKW15, WWZ16}
%Let $O=\{o_1, \cdots, o_n\}$ and let $O'=\{o_1,\ldots, o_n'\}$. 
%Let $I$ be set which is disjoint from $O\union O'$. 
Let $\R$ be a reactive system design and let $D$ be a safety specification, 
both over $(I, O)$. Then a safety shield for $\R$ and $D$ is a reactive system 
$\Shield$ over $(I, O')$ satisfying:
\begin{itemize}
\item $\R\circ\Shield\models D[O/O']$. 
\item For all input traces $\alpha=\alpha_0\alpha_1\cdots$ 
$(\R\circ\Shield)(\alpha_i)$ ``deviates'' from $\R(\alpha_i)$ as seldom as possible. 
\end{itemize}
%i.~e.~all the composed behaviours of $\R$ and $\Shield$ satisfy $D[O/O']$, 
\end{definition}
The word ``deviates'' assumes different meaning in the literature. 
For example, in \cite{BKKW15} Bloem et.~al.~proposed \emph{$K$-shield} which can 
disagree with the design for at most $K$ consecutive steps 
provided the design recovers immediately after an error and does not 
violate the specification for next $K$ cycles. 
Since the shield is allowed a window of $K$ cycles to deviate from 
the design output in the event of an output error by the design 
it is not suited to handle burst errors. 
In \cite{WWZ16} Wu et.~al.~proposed a {\em burst error shield} which 
is resilient to burst error and matches the design output 
whenever it meets the specification. 
However, this strategy of matching the design output until a violation occurs makes 
the synthesis algorithm ``non-conservative'' in the sense that 
it may fail to generate a shield even if specification is realizable. 

To overcome these issues we propose 
\emph{conservative shield synthesis}. 
The conservative shield synthesis differs from $K$-shield synthesis 
in two key respects: it does not impose any restriction on the design output 
and it can handle burst errors. 
It also differs from Wu et.~al.~as, unlike them, we will always synthesize a shield 
whenever the specification is realizable. 
%that will always try to match the design output as {\em often} as possible 
%shields which are able to handle the burst errors. 

Shield synthesis criteria can be specified using hard and soft requirements. 
\dcsynth can then synthesize the desired shield. We give examples of variants 
of $K$-shield and Burst error shield, and call the ``conservative''.
%We now focus on how to synthesize conservative $K$-shield and 
%conservative burst error shield for a specification \verb|REQ| over $(I,O)$. 
\begin{itemize}
\item Conservative $K$-shield.
\begin{itemize}
\item Input: $I\union O$. Output: $O'$
\item Hard requirement: \verb|REQ[O/O']&&[]([[|$\vee_{o\in O}$\verb|(o|$\neq$\verb|o')]]=>slen<k)|. 
\item Soft requirement: None
\end{itemize}  
\item Conservative burst error shield.
\begin{itemize}
\item Input: $I\union O$. Output: $O'$
\item Hard requirement: \verb|REQ[O/O']|
\item Soft requirement: $\wedge_{o\in O}\verb|(true ^ <o=o'>)|$ with all of formulas assigned same priority. 
\end{itemize}  

%\hspace{1cm} 
We emphasize the importance soft requirements here, 
%Apart from handling burst errors, 
it  forces a conservative burst error shield to 
try and match the design output as often as possible. 
%and deviates from it only if it is necessary. 
%Our soft requirement plays a key role here for us do this in \dcsynth. 
%Let $D^{Soft'} = \forall 1\leq i\leq n: o_i=o_i'$ 
%with same priority assigned to each one.
This is because when all the soft requirements are assigned same priority 
\dcsynth will try to synthesize a controller selecting an output 
which meets the hard requirement as well as a maximal set of soft requirements at every step. 
%So, with $D^{Soft'}$ as the soft requirement, 
%the shield synthesized by \dcsynth will try to match as many outputs of $\R$ as possible 
%at every step ensuring minimal deviation. % from the output of $\R$. 
%We note that though similar in spirit 
%a conservative safety shield is not exactly a burst error shield.
\end{itemize}.

We have rerun the experiments in \cite{WWZ16} in our framework, 
and the results are as tabulated in Table.~\ref{tab:dfaBasedComparision}. 
For the sake of comparison we use the input files of Wu et.~al.~\cite{Wu16a} 
as inputs to \dcsynth\footnote{Automata as formula}. As the table suggests, 
in most of the cases the shield that we synthesize compares favorably 
with the corresponding shield synthesized in \cite{BKKW15} and \cite{WWZ16} 
both in terms of size and time taken for the synthesis. 
For instance, for the guarantee AMBA G5+6+9e64+10 our tool 
synthesize a shield significantly faster and 
with smaller no.~of states than the existing tools\cite{BKKW15,WWZ16}.

\begin{table}
\caption {Comparison of K-Shield and Burst error shield with Conservative safety shield. 
Columns under $K$-shield and Burst error shield are taken from Wu et.~al.~\cite{WWZ16} 
and reproduced here for comparision.}
\label{tab:dfaBasedComparision}
\centering
\resizebox{\columnwidth}{!}{%
%\begin{threeparttable}[b]
%\begin{center}
\begin{tabular}{||@{\quad}c@{\quad}|@{\quad}c@{\quad}|@{\quad}c@{\quad}|@{\quad}c@{\quad}|@{\quad}c@{\quad}|@{\quad}c@{\quad}|@{\quad}c@{\quad}|@{\quad}c@{\quad}|} 
\hline
\multirow{2}{*}{Guarantees}&\multirow{2}{*}{Monitor states}&\multicolumn{2}{l|}{K-Shield}&\multicolumn{2}{l|}{Burst error shield}&\multicolumn{2}{l|}{Conservative burst error shield}\\\cline{3-4}\cline{5-6}\cline{7-8} 
&&states&time&states&time&states&time\\\hline\hline
Toyota Powertrain & 23  & 38        & 0.2       & 38 & 0.3 & 9 & 0.7 \\\hline
Traffic light         & 4  & 7        & 0.1       &7 & 0.2 & 4 & 0.007\\\hline
$F_{64}p$         & 67  & 67   & 0.7       & 67 & 0.5 & 67 & 0.002\\\hline
$F_{256}p$         & 259  & 259    & 46.9 & 259 & 10.5 & 259 & 0.01 \\\hline
$F_{512}p$         & 515  & 515 & 509.1 & 515  & 54.4  & 515 & 0.07 \\\hline
G($\neg$ q) $\vee$  $F_{64}$(q $\wedge$ $F_{64}p$)    &67  & 67   & 0.8 & 67 & 0.6 & 67 & 0.007\\\hline
G($\neg$ q) $\vee$  $F_{256}$(q $\wedge$ $F_{256}p$) &259  & 259 & 46.2 & 259 & 10.7 & 259 & 0.04\\\hline
G($\neg$ q) $\vee$  $F_{512}$(q $\wedge$ $F_{512}p$) & 515  & 515 & 571.7 & 515 & 54.5 & 515 & 0.1\\\hline
G(q $\wedge$ $\neg$ r $\rightarrow$ ($\neg$ r $\cup_{4}$ (p $\wedge$ $\neg$ r))) & 6  & 15 & 0.1 & 145 & 0.1 & 6 & 0.004\\\hline
G(q $\wedge$ $\neg$ r $\rightarrow$ ($\neg$ r $\cup_{8}$ (p $\wedge$ $\neg$ r))) & 10  & 109  & 0.2 & 5519 & 4.5 & 10 & 0.005\\\hline
G(q $\wedge$ $\neg$ r $\rightarrow$ ($\neg$ r $\cup_{12}$ (p $\wedge$ $\neg$ r))) & 14  & 753 & 6.3 & 27338 & 1414.5 & 14 & 0.006\\\hline \hline
AMBA G1+2+3         & 12  & 22  & 0.1   & 22 & 0.1 & 7 & 0.008\\\hline
AMBA G1+2+4    & 8   & 61 & 6.3 & 78 & 2.2 & 8 & 0.6\\\hline
AMBA G1+3+4      & 15   & 231  & 55.6 & 640 & 97.6 & 14 & 0.4 \\\hline
AMBA G1+2+3+5    & 18   & 370 & 191.8  & 1405 & 61.8 & 17 & 0.05 \\\hline
AMBA G1+2+4+5      & 12   & 101  & 3992.9 & 253 & 472.9 & 12 & 3.2 \\\hline
AMBA G4+5+6      & 26   & 252 & 117.9 & 205 & 26.4 & 18 & 0.6\\\hline
AMBA G5+6+10     & 31   & 329 & 9.8 & 396 & 31.4 & 27 & 2.6 \\\hline
AMBA G5+6+9e4+10 & 50   & 455 & 17.6 & 804 & 42.1 & 46 & 5.2\\\hline
AMBA G5+6+9e8+10 & 68   & 739 & 34.9 & 1349 & 86.8 & 64 & 7.6\\\hline
AMBA G5+6+9e16+10 & 104   & 1293 & 74.7 & 2420 & 189.7 & 100 & 12.5\\\hline
AMBA G5+6+9e64+10 & 320   & 4648 & 1080.8 & 9174 & 2182.5 & 316 & 40.9\\\hline
AMBA G8+9e4+10   & 48   & 204 & 7.0 & 254 & 6.1 & 48 & 0.3\\\hline
AMBA G8+9e8+10   & 84   & 422 & 22.5 & 685 & 33.7 & 84 & 0.5 \\\hline 
AMBA G8+9e16+10   & 156   & 830 & 83.7 & 1736 & 103.1 & 156 & 0.9\\\hline 
AMBA G8+9e64+10   & 588   & 3278 & 2274.2 & 7859 & 2271.5 & 588 & 3.3 \\\hline 
\end{tabular}
%\end{center}%
}
\end{table}

\section{Quantitative Latency Measurement}
\label{section:quantitative}

Soft requirements are often used as directives to the synthesis algorithm which impact the ``latency'' of the controller. We give a 
notation to allow users to specify what is \emph{latency}. Model checking technique, implemented in tool CTLDC \cite{Pan05}, can
then measure the worst case latency.

For latency measurement, user must specify a \qddc\/ formula $D^{p}$ characterizing execution fragments of interest. 
For example the \qddc formula $D^{p}$ = \verb#[[req]]&&(scount ack <3)# specifies fragments of execution with 
request continuously {\em true} 
but with less than 3 acknowledgments (exact syntax is explained in \S\ref{section:dcsynth-spec}).
The latency goal $MAXLEN(D^{p},M)$
computes $\sup \{e-b \mid \rho[b,e] \models D^{p}, \rho \in Exec(M) \}$, i.~e.~it computes the length of
the longest interval satisfying $D^{p}$ within the executions of $M$. 
For example $MAXLEN(D^{p},Arb)$ specifies the worst case response time of the arbiter $Arb$  to get three acknowledgments. 
Tool CTLDC, which like \DCSYNTH\/ is member of DCTOOLS suite of tools, provides efficient computation of $MAXLEN$ by symbolic search for longest 
paths as formulated in \cite{Pan05}.

\oomit{
Given a controlled system $M ~=~ A(Assumptions) \times Controller$ as safety automaton $M$, and
a duration calculus formula $D$, aim is to compute $MAXLEN(D,M)$. We assume that all states of $M$ are reachable.
We first construct NFA $M'$ by turning every accepting state of $M$
as initial state. Clearly $M'$ accepts all fragments of behaviors of $M$. Let $N = M' \times A(D)$. Then,
$N$ accepts all fragments of behaviors of $M$ which satisfy $D$. Now we find the length of the longest accepting path in $N$. Campos \emph{et al} have given symbolic BDD-based algorithm for computing longest and shortest such paths \cite{CCMMH94}. We adapt and implement this in tool DCSynth to compute  the
values of $MAXLEN(D,M)$ as well as $MINLEN(D,M)$. 
This method was previously proposed in \cite{Pan05} where it was shown to be efficient. 
}

Table.~\ref{tab:perfMeasure2} gives worst case latency measurements carried out using tool CTLDC for various controllers synthesized using \DCSYNTH.
The results illustrate the impact of soft goals on controller behaviour as well as controller latency under various scenarios. 
For example, $Arb^{soft}(6,2)$ ``tries'' to give acknowledgement within 2 cycles with 
higher priority assigned to higher numbered cell (see the description in \S\ref{section:motivation}). 
Note that response time is {\em one} more than that in the column \textsl{Computed Response} in the Table. 
The worst case latency measurement shows that \verb#req6# has response time of 1 cycle 
whereas \verb#req5# has response time of 2 cycles. 
For all other cells  the response time is $\infty$ since cells 6 and 5 can consume all the cycles.
Note that when \verb#req6# is absent throughout 
the response time of cell 4 changes from 
$\infty$ to 3 cycles as shown in the $4^{th}$ row of the Table.~\ref{tab:perfMeasure2}. 
This points to the {\em robustness} of
the synthesized controller.

\begin {table}[!h]
\caption {Worst Case Latency Analysis using CTLDC using MAXLEN(Response Formula) computation}
\label{tab:perfMeasure2}
\begin{center}
	\begin{tabular}	{|c|c|c|c|}
	\hline
		Sr.No & Example & Response Formula & Computed Response \\
%		& & Using MAXLEN & (In Cycle)\\
	 \hline
		1 &  $Arb^{soft}(6,2)$ & ([[req6]]\&\&([[!ack6]])) & 1 \\
%		& & \hrulefill & \hrulefill \\
%	    & & [[req6]] \&\& ((scount ack6 $<$ 3)) & 5 \\
	 \hline
	    2 & $Arb^{soft}(6,2)$ & ([[req5]]\&\&([[!ack5]])) & 2 \\
%	    & & \hrulefill & \hrulefill \\
%	    & & [[req5]] \&\& ((scount ack5 $<$ 3)) & 8\\
	 \hline
	    3 & $Arb^{soft}(6,2)$ & ([[$req_i$]]\&\&([[!$ack_i$]])) & $\infty$ \\
%	    & & \hrulefill & \hrulefill \\
	    & for $1 \leq i \leq 4$ & & $\infty$ \\
%	    & for $1 \leq i \leq 4$ & [[$req_i$]] \&\& ((scount $ack_i$ $<$ 3)) & $\infty$ \\
	    
	 \hline
	    4 & $Arb^{soft}(6,2)$ & ([[req4 \&\& !req6]] \&\& ([[!ack4]])) & 2 \\
	 \hline
	    5 & $Arb^{soft}(5,3)$ & ([[req5]]\&\&([[!ack5]])) & 2 \\
%	    & & \hrulefill & \hrulefill \\
%	    & & [[req5]] \&\& ((scount ack5 $<$ 3)) & 8 \\
	 \hline
	    6 & $Arb^{soft}(5,3)$ & ([[req4]]\&\&([[!ack4]])) & 3 \\
%	    & & \hrulefill & \hrulefill \\
%	    & & [[req4]] \&\& ((scount ack4 $<$ 3)) & 11 \\
	 \hline
	    7 & $Arb^{soft}(5,3)$ &  ([[req3]]\&\&([[!ack3]])) & 4 \\
%	    & & \hrulefill & \hrulefill \\
%	    & & [[req3]] \&\& ((scount ack3 $<$ 3)) & 14\\
	 \hline
		8 & MINEPUMP(MPV1) & [[AssumptionOk \&\& HH2O]] & 4 \\
	 \hline
	    9 & MINEPUMP(MPV2) & [[AssumptionOk \&\& HH2O]] & 7 \\
	 \hline
	    10 & MINEPUMP(MPV3) & [[AssumptionOk \&\& HH2O]] & 8 \\
	 \hline
	\end{tabular}
\end{center}
\end{table}
For the $MINEPUMP$ case study rows 8,9,10  give the maximum amount of time (in cycles) 
for which the water level can remain high (indicated by the variable {\em HH2O}) 
without violating the assumptions (indicated by {\em AssumptionOk}). Here, $MINEPUMP(req)$
denotes $MINEPUMP$ specification with soft requirement $req$ as given in  Appendix \ref{section:minepumpexample}.
For example, soft requirement MPV3 is \verb#!PumpOn# which tries to keep pump {\em off} 
as much as possible where as
soft requirement MPV1 is \verb#PumpOn# which tries to keep pump {\em on} as much as possible. 
As a result $MINEPUMP(MPV1)$ gets rid of water in 4 cycles compared to 8 cycles for $MINEPUMP(MPV3)$.

%pandya

%\input{Discussion}
%paritosh
\bibliographystyle{plain}
\bibliography{rmmRef}
\clearpage
\appendix
\section{Logic \qddc}
\label{section:qddc}

Let $\Sigma$ be a finite non empty set of propositional variables. 
A \emph{word} $\sigma$ over $\Sigma$ is a finite sequence of the form 
$P_0\cdots P_n$ where $P_i\subseteq\Sigma$ for each $i\in\{0,\ldots,n\}$. 
%For a word $\sigma=a_0\cdots a_n$ 
Let $\len{\sigma}=n+1$, 
$\dom{\sigma}=\{0,\ldots,n\}$ and $\forall i\in\dom{\sigma}:\sigma(i)=P_i$.  

The syntax of a \emph{propositional formula} over $\Sigma$ is given by:
\[
\varphi := \0\ |\ \1\ |\ p\in\Sigma\ |\ \varphi\wedge\varphi\ |\ \varphi\Or\varphi\ |\ \neg\varphi, 
\]
and operators such as $\Rightarrow$ and $\iff$ are defined as usual. 
Let $\Omega_\Sigma$ be the set of all propositional formulas over $\Sigma$. 

%Let $\sigma=P_0\cdots P_n$ be a word and $\varphi\in\Omega_\Sigma$. 
Let $i\in\dom{\sigma}$. 
Then the satisfaction relation $\sigma,i\models\varphi$ is defined inductively as follows:
%$\sigma, i \models \1$; $\sigma, i \models p$ iff $p\in\sigma(i)$; 
%$\sigma, i \models \neg p$ iff $\sigma,i\not\models p$, 
\[
\begin{array}{lcl}
\sigma, i \models \1, & &\\
\sigma, i \models p & \mathrm{\ iff \ } & p\in\sigma(i),\\
\sigma, i \models \neg p & \mathrm{\ iff \ } & \sigma,i\not\models p,\\
\end{array}
\]
and the satisfaction relation for the rest of the 
boolean combinations defined in a natural way.

The syntax of a \qddc formula over $\Sigma$ is given by: 
\[
\begin{array}{lc}
D:= &\ang{\varphi}\ |\ \sq{\varphi}\ |\ \dsq{\varphi}\ |\ \dcurly{\varphi}\ |\ D\ \verb|^|\ D\ |\ \neg D\ |\ D\Or D\ |\\ 
&D\wedge D\ |\ D^*\ |\ \exists p.\ D\ |\ \forall p.\ D\ |\\ 
&slen \bowtie c\ |\ scount\ \varphi \bowtie c\ |\ sdur\ \varphi \bowtie c, 
\end{array} 
\]
where $\varphi\in\Omega_\Sigma$, $p\in\Sigma$, 
$c\in\nat$ and $\bowtie\in\{<,\leq,=,\geq,>\}$. 

An \emph{interval} over a word $\sigma$ is of the form $[b,e]$ 
where $b,e\in\dom{\sigma}$ and $b\leq e$. 
An interval $[b_1,e_1]$ is a sub interval of $[b,e]$ 
if $b\leq b_1$ and $e_1\leq e$. 
Let $\intv{\sigma}$ be the set of all intervals over $\sigma$.

%Let $\Sigma$ be a set of propositional variables. 
%and let $\Omega$ be the set of all propostional formulas over $\Sigma$. 
Let $\sigma$ be a word over $\Sigma$ and let $[b,e]\in\intv{\sigma}$ be an interval. 
Then the satisfaction relation of a \qddc formula $D$ over 
$\Sigma$, written $\sigma,[b,e]\models D$, is defined inductively as follows:
\[
\begin{array}{lcl}
\sigma, [b,e]\models\ang{\varphi} & \mathrm{\ iff \ } & \sigma,b\models \varphi,\\
\sigma, [b,e]\models\sq{\varphi} & \mathrm{\ iff \ } & \forall b\leq i<e:\sigma,i\models \varphi,\\
\sigma, [b,e]\models\dsq{\varphi} & \mathrm{\ iff \ } & \forall b\leq i\leq e:\sigma,i\models \varphi,\\
\sigma, [b,e]\models\dcurly{\varphi} & \mathrm{\ iff \ } & e=b+1 \mbox{ and }\sigma,b\models \varphi,\\
\sigma, [b,e]\models\neg D & \mathrm{\ iff \ } & \sigma,[b,e]\not\models D,\\
\sigma, [b,e]\models D_1\Or D_2 & \mathrm{\ iff \ } & \sigma, [b,e]\models D_1\mbox{ or }\sigma,[b,e]\models D_2,\\
\sigma, [b,e]\models D_1\wedge D_2 & \mathrm{\ iff \ } & \sigma, [b,e]\models D_1\mbox{ and }\sigma,[b,e]\models D_2,\\
\sigma, [b,e]\models D_1\verb|^| D_2 & \mathrm{\ iff \ } & \exists b\leq i\leq e:\sigma, [b,i]\models D_1\mbox{ and }\\
&&\sigma,[i,e]\models D_2.\\
\end{array}
\]
We call word $\sigma'$ a $p$-variant, $p\in\Sigma$, of a word $\sigma$ 
if $\forall i\in\dom{\sigma},\forall q\neq p:\sigma'(i)(q)=\sigma(i)(q)$. 
Then $\sigma,[b,e]\models\exists p.~D\iff\sigma',[b,e]\models D$ for some 
$p$-variant $\sigma'$ of $\sigma$ and, 
$\sigma,[b,e]\models\forall p.~D\iff\sigma,[b,e]\not\models\exists p.~\neg D$. 
We define $\sigma\models D$ iff $\sigma,[0,\len{\sigma}]\models D$.

\begin{example}
Let $\Sigma=\{p,q\}$ and let 
$\sigma=P_0\cdots P_7$ be such that $\forall 0\leq i<7:P_i=\{p\}$ and $P_7=\{q\}$. 
Then $\sigma, [0,7]\models\sq{p}$ but not $\sigma, [0,7]\models\dsq{p}$ 
as $p\not\in P_7$.   
\end{example}

\begin{example}
Let $\Sigma=\{p,q,r\}$ and let 
$\sigma=P_0\cdots P_{10}$ be such that $\forall 0\leq i<4:P_i=\{p\}$, 
$\forall 4\leq i<8:P_i=\{p,q,r\}$ and $\forall 8\leq i\leq 10:P_i=\{q,r\}$.  
Then 
\[
\sigma, [0,10]\models \sq{p}\verb|^|\dsq{\neg p\wedge r}
\]
because for $i\in\{8,9,10\}$ 
the condition $\exists 0\leq i\leq 10:\sigma, [0,i]\models\sq{p}$ and 
$\sigma,[i,10]\models\dsq{\neg p\wedge r}$ 
is met. But $\sigma, [0,7]\not\models\sq{p}\verb|^|\dsq{\neg p\wedge r}$ as 
$\neg\exists 0\leq i\leq 7:\sigma, [0,i]\models\sq{p}\mbox{ and }\sigma,[i,7]\models\dsq{\neg p\wedge r}$. 
%$i\in\{0,\ldots,7\}$ which will meet the condition. 
\end{example}

Entities \slen, \scount, and \sdur are called \emph{terms}. 
The term \slen gives the length of the interval in which it is 
measured, $\scount\ \varphi$ where $\varphi\in\Omega_\Sigma$, counts 
the number of positions including the last point 
in the interval under consideration where $\varphi$ holds, and    
$\sdur\ \varphi$ gives the number of positions excluding the last point 
in the interval where $\varphi$ holds. 
Formally, for $\varphi\in\Omega_\Sigma$ we have 
%$\slen(\sigma, [b,e])=e-b$, 
%$\scount(\sigma,\varphi,[b,e])=\sum_{i=b}^{i=e}\left\{\begin{array}{ll}
%					1,&\mbox{if }\sigma,i\models\varphi,\\
%					0,&\mbox{otherwise.}
%					\end{array}\right\}$ and 
%$\sdur(\sigma,\varphi,[b,e])=\sum_{i=b}^{i=e-1}\left\{\begin{array}{ll}
%					1,&\mbox{if }\sigma,i\models\varphi,\\
%					0,&\mbox{otherwise.}
%					\end{array}\right\}$
\[
\begin{array}{lcl}
\slen(\sigma, [b,e])&=&e-b,\\
\scount(\sigma,\varphi,[b,e])&=&\sum_{i=b}^{i=e}\left\{\begin{array}{ll}
					1,&\mbox{if }\sigma,i\models\varphi,\\
					0,&\mbox{otherwise.}
					\end{array}\right\}\\
\sdur(\sigma,\varphi,[b,e])&=&\sum_{i=b}^{i=e-1}\left\{\begin{array}{ll}
					1,&\mbox{if }\sigma,i\models\varphi,\\
					0,&\mbox{otherwise.}
					\end{array}\right\}
\end{array}
\]
In addition we also use the following derived constructs: 
$\sigma, [b,e]\models\pt$ iff $b=e$; $\sigma, [b,e]\models\ext$ iff $b<e$; 
$\sigma, [b,e]\models \Diamond D$ iff $\true\verb|^|D\verb|^|\true$ and 
$\sigma, [b,e]\models \Box D$ iff $\sigma, [b,e]\not\models\Diamond\neg D$. 

A \emph{formula automaton} for a \qddc formula $D$ is a 
\emph{deterministic finite state automaton} which accepts precisely language 
$L=\{\sigma\ |\ \sigma\models D\}$.
%It is known from \cite{Pan01} that: 

\begin{theorem}
\label{theorem:formula-automaton}\cite{Pan01a}
For every \qddc formula $D$ over $\Sigma$ we can construct a DFA $\A(D)$ 
for $D$ such $L(D)=L(\A(D))$. 
%The size of $\A(D)$ is non elementary in the size of $D$ in the worst case.
\end{theorem}

\clearpage

\section{The Algorithm Implementation}
\label{subsection:algorithms}
The example shows the explicit state representation of the 
state space to produce the controller for demonstration purpose.

The algorithms are actually designed to work on symbolic data structure to 
represent the transistion function for each automaton.
We use Multi-Terminal BDD(MTBDD) to represent the all the automaton. 

The psuedocode of our algorithm is given in the following section.

\begin{algorithm}
\label{algo:synthesis}
\textbf{SYNTHESIZE:}

\textbf{Input}: $S=( I,O,D^h,\wedge_{i=1}^{i=k}(w_i\iff D_i^s), \langle P_1,\cdots,P_l\rangle)$

\textbf{Output}: Controller for S.

1. $A^{mon}$=GenMonitorAutomaton(S) 

\hspace{1 cm}//Generates prefix closed language equivalent safety automaton

2. $A^{mpnc}$=GenMPNC($A^{mon}$, I, O) 

\hspace{1 cm}//Generates MPNC from $A^{mon}$ and Input-output partitioning

3. IF initial state of $A^{mpnc}$ is NOT an accepting state THEN

	\hspace{1 cm}	S is unrealizable,
		generate a counter example tree
		
	\hspace{0.5 cm} ELSE
	
	\hspace{1 cm}Specification is realizable, GOTO step 4.
	
4. $A^{lodc}$ = GenLODC($A^{mpnc}$, $\langle P_1$, $\ldots$, $P_l \rangle$)

	\hspace{1 cm}//Determinizes the MPNC with respect to Soft Requirements.
		
5. Encode $A^{lodc}$ in an implementation language.

\end{algorithm}

\vspace{1cm}
The monitor automaton $A^{mon}$ is obtained by GenMonitorAutomaton(),
based on the procedure implemented in a tool DCVALID.
 
The algorithm for construction of $A^{mpnc}$ from $A^{mon}$,
is implemented by the function GenMPNC(). To illustrate this function,
we first define the function $C_{step}$: S $\times$ $2^S$ $\rightarrow$ \{1, 0\} as follows: 

\hspace{0.5 cm}$C_{step}$(s, G) = 1 if $\forall$ i, $\exists$ o : $\delta(s, (i,o))$ $\in$ G

\hspace{1.5 cm}else $C_{step}$(s, G) = 0.

\hspace{0.5 cm}where i $\in$ I and o $\in$ (O $\cup$ W) and s $\in$ S.

\begin{algorithm}
\label{algo:genmpnc}
\textbf{GenMPNC:}

\textbf{Input}: $A^{mon}$, I, O, W

\textbf{Output}: $A^{mpnc}$.

S = set of states in $A^{mon}$, 

F = set of accepting states in $A^{mon}$

$\delta$: S $\times$ $(I \cup O \cup W)$ $\rightarrow$ S be the trasition function in $A^{mon}$.

$\mathcal{V}$: S $\rightarrow$ \{1, 0\} be a value function over S

Initialize $\mathcal{V}$(s)=1 $\forall$s $\in$ G, otherwise $\mathcal{V}$(s) = 0

%LET G $\subseteq$ S be a set of good(accepting) states.

SET G = F

DO

	\hspace{0.5 cm}Pre\_$\mathcal{V}$ = $\mathcal{V}$
	
	\hspace{0.5 cm}FOR each s $\in$ G do
	
		\hspace{1 cm}IF $C_{step}$(s, G) = 0 then 
		
			 \hspace{1.5 cm}$\mathcal{V}$(s) = 0
			 
		\hspace{1.5 cm}G = G $-$ s
			 
WHILE (Pre\_$\mathcal{V}$ $\neq$ $\mathcal{V}$ )

$A^{mpnc}$ = Created from $A^{mon}$ by keeping only the states s $\in$ G and transitions (s, t) s.t. s $\in$ G and t $\in$ G. 

\end{algorithm}

\vspace{0.5 cm}

Now, we give construction of $A^{lodc}$ from $A^{mpnc}$, 
which is implemented by the function GenLODC(). GenLODC()
determinizes the automaton such that for any input a unique
output can be selected.

We define the function \emph{evaluateSoftReq}($\langle$ $P_1$,..,$P_l$ $\rangle$, 
input output valuation), which takes
list of soft requirements and input-output valuation as input and returns the weighted 
value of the soft requirements being satisfied by the valuation. 

We also define \emph{lookupTable:I $\rightarrow$ (O $\times$ S $\times$ Integer)}, 
which contains for each input, the output value and the next state, that maximizes 
the satisfaction of soft requirements.
Similarly a function \emph{initLookup: lookupTable $\times$ val $\rightarrow$ lookupTable}
 initializes the lookupTable to some minimal value val for each input valuation..

\begin{algorithm}
\label{algo:genlodc}
\textbf{GenLODC:}

\textbf{Input}: $A^{mpnc}$, I, O, $\langle$ $P_1$,..,$P_l$ $\rangle$

\textbf{Output}: $A^{lodc}$.

S = set of states in $A^{mpnc}$, 

%r = a unique reject state in $A^{mpnc}$
%MPNC has a unique reject state '.'

%LET lookupTable:I $\rightarrow$ (O $\times$ S $\times$ Integer}, contains for each input, the optimal value, optimal output, and the next state.

initLookup(lookupTable, -1) initializes the lookupTable by the -1 for each input.

FOR every state s $\in$ S DO

\hspace{0.5 cm} FOR every valuation (i,o,w) of (I $\cup$ O $\cup$ W)

	\hspace{1 cm} val = evaluateSoftReq($\langle$ $P_1$,..,$P_l$ $\rangle$, (i,o,w))
	
	\hspace{1 cm} IF val $>$ lookupTable(i) THEN
				lookupTable(i) = \{val, o, next\_state\}
				
Create the Automaton $A^{lodc}$ with updated transitions with valuation for each 
input and the corresponding output valuation given in lookupTable for each state.

\end{algorithm}

\subsubsection{Complexity Results:}

The algorithms work directly on this symbolic representaion,
including function $C_{step}$ used inside GenMPNC.
For the algorithn GenMPNC, the worst case complexity is 
$\mathcal{O}(N^2.|BDD|)$,
where N is the number of states in monitor automaton $A^{mon}$. 
This can be derived from the fact that the maximum number of 
iterations to reach a fix-point is ($N-1$),
and in each iteration there can be $\mathcal{O}(N)$ step for
each state which are marked as good.
Each such state may requires $\mathcal{O} (|BDD|)$ steps 
to determine whether the state is winning or not 
(determined by function $C_{step}$). $|BDD|$ represents the 
size of the MTBDD datastructure in terms of the number of BDD nodes.

The function $C_{step}$ is an important function as it is
the core function used to find the winning region.
We implement this function over MTBDD data structure 
\emph{without actually creating a game graph}. 

We give the outline of our algorithm as follows:

We assume that safety monitor automaton $A^{mon}$ is given as a 
dfa $M=(Q,\Sigma,\delta,G)$ where $G \subseteq Q$  is the set of accepting states.
$\Sigma = 2^{(I \cup O \cup W)}$ is the alphabet and 
$\delta: Q \times \Sigma \rightarrow Q$ with $Q-G$ 
being the reject states.
We assume that $\delta$ is encoded as MTBDD as in MONA.

In MTBDD the bdd nodes can be categorized as \emph{internal nodes} or the \emph{terminal node}.
The terminal node represent the destination state of the transition. The internal 
nodes represent the decision on some variable $v \in (I \cup O \cup W)$.

Our algorithm start by labelling each terminal node with a value and then
propagating this value to the bdd node corresponding to 
source state to see whether the source state belong
to the winning region or not. 
\begin{enumerate}

\item We starts by labeling each terminal node. Every terminal node 
is labelled as 1 if it belongs to accepting states, otherwise it is labelled as 0.

\item then we label the internal bdd nodes whose successors are already labelled as follows:
    if the internal node represents the decision on an input variable then its lable is minimum of its successors. Otherwise (representing decision on output or indicatior variable), the internal node is labeled by maximum of its sucessor.
    
\item step 2 is performed recursively until we get the bdd node corresponding to 
sources (starting) state labeled with 0 or 1. If bdd node for source state is
labelled as 1 then it is inside the winning region based on current labeling
of terminal nodes. If bdd node for source state is labelled as 0 then it's not
in the winning region and can be removed during construction of MPNC.

\end{enumerate}

\begin{figure}
\label{fig:mtbddex1}
\begin{center}
\resizebox{7cm}{!}{
\includegraphics{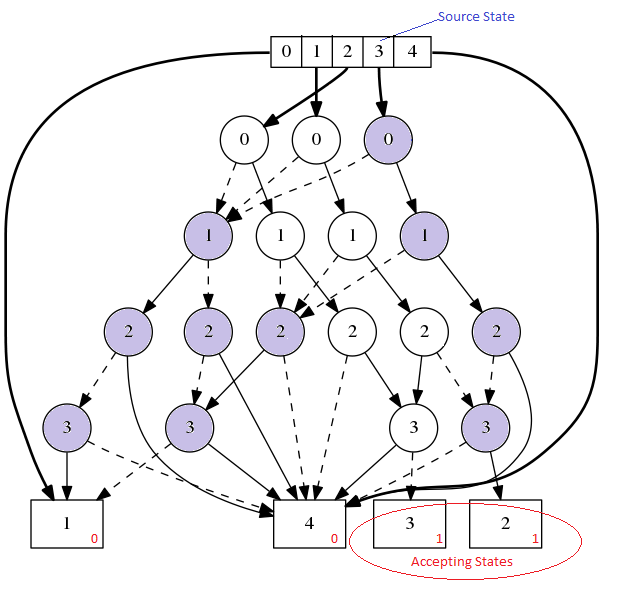}
}
\end{center}
\caption{MPNC Computation over MTBDD for state 3: BDD nodes colored purple to be evaluated}
\end{figure}

\begin{figure}
\label{fig:mtbddex2}
\begin{center}
\resizebox{7cm}{!}{
\includegraphics{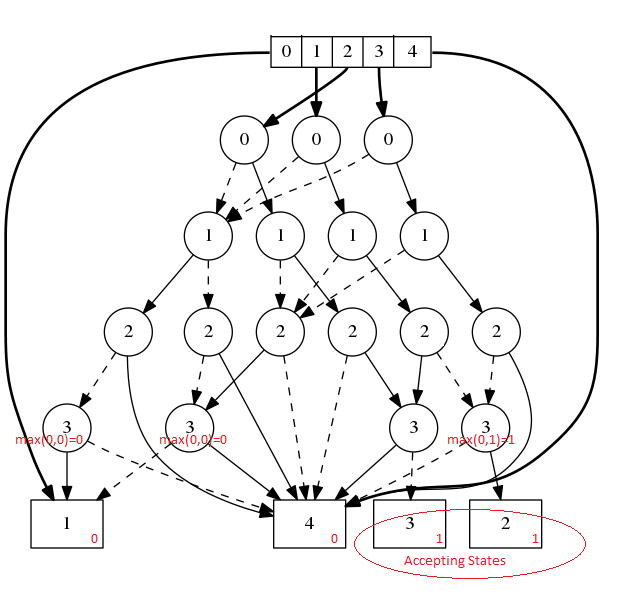}
}
\end{center}
\caption{MPNC Computation over MTBDD for variable index 3}
\end{figure}

\begin{figure}
\label{fig:mtbddex3}
\begin{center}
\resizebox{7cm}{!}{
\includegraphics{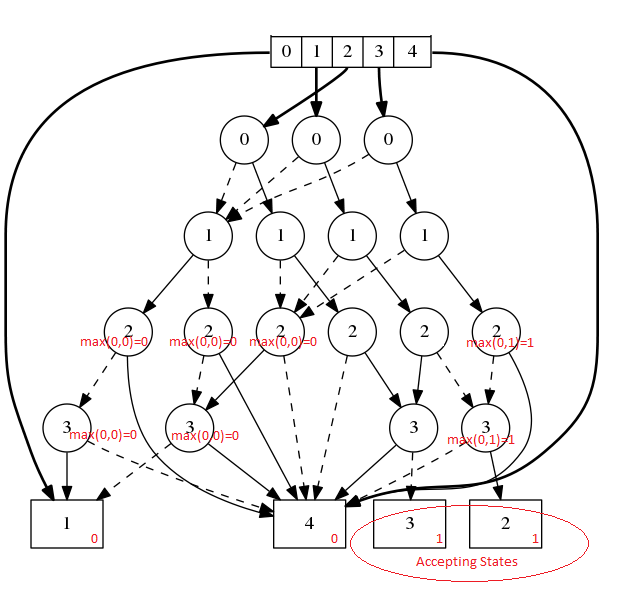}
}
\end{center}
\caption{MPNC Computation over MTBDD for variable index 2}
\end{figure}

\begin{figure}
\label{fig:mtbddex4}
\begin{center}
\resizebox{7cm}{!}{
\includegraphics{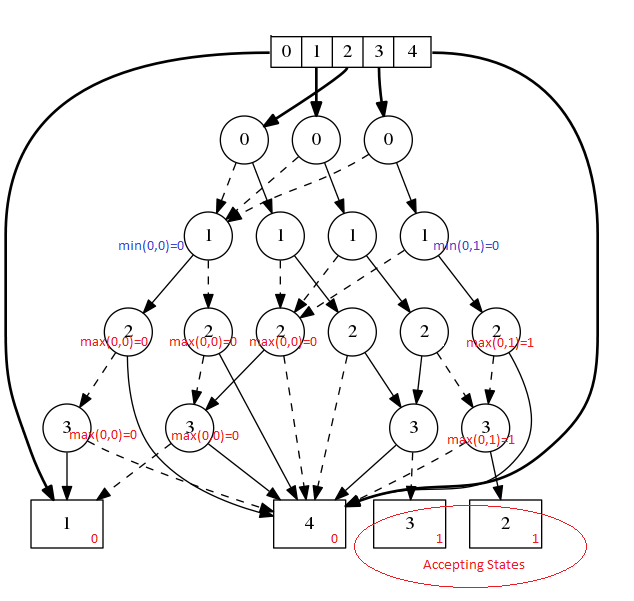}
}
\end{center}
\caption{MPNC Computation over MTBDD for variable index 1}
\end{figure}

\begin{figure}
\label{fig:mtbddex5}
\begin{center}
\resizebox{7cm}{!}{
\includegraphics{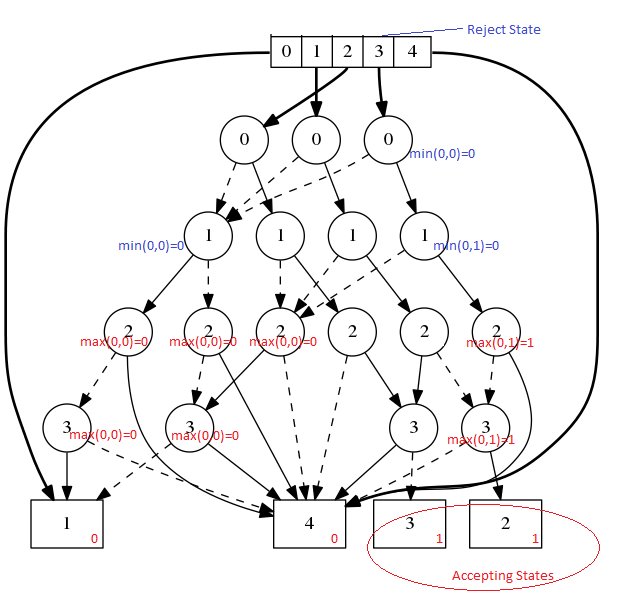}
}
\end{center}
\caption{MPNC Computation over MTBDD for variable index 0}
\end{figure}
The example computation of $C_{step}$ function of the MTBDD
for source state 3 and starting with the winning region states \{3, 4\} is shown in the 
figures from ~\ref{fig:mtbddex1} to ~\ref{fig:mtbddex5}. 
In the MTBDD there are 4 variables indexed as 0,1,2 and 3. 
 The variable indexed 0,1 are inputs and 2, 3 are outputs.
All internal nodes are represented by circle and encircled number is the decision variable.
All terminal nodes are represented by rectangle and number inside that is the destination state number.
The list on the top corresponds to the state numbered 0 to 4.

Similarly the worst case complexity for GenLODC 
could be calculated as $\mathcal{O}(N.l.(2^{|I \cup O \cup W|}))$, 
where N is the number of states in MPNC, l is the number of soft requirements.
%and $F$ is average size of soft requirement formula. 
This can be derived from the fact that maximum number of states in LODC can
be equal to MPNC in case MPNC itself is a deterministic automaton.
For each state in LODC we have to compute its locally optimal output,
which depends on the number of soft requirements to be computed.

Although in worst case $|BDD|$ can be $\mathcal{O}(2^{|I \cup O \cup W|})$,
but in most of the practical examples the size of MTBDD is much smaller,
and therefore the tool performs much better that the worst case estimation.

In Section ~\ref{subsection:lodcoptimization} we give an
algorithm for GenLODC to exploit the shared bdd nodes in MTBDD.
The results show the considerable improvement over the naive
explicit path enumeration based algorithm.

\subsection{LODC Optimization}
\label{subsection:lodcoptimization}
We have also developed an algorithm to efficiently compute the LODC from MPNC.
Following section gives the brief overview of algorithm.
%%%%%%%%%%%
\subsubsection{Outline of Optimization Algorithm}

We assume that MPNC is given as a dfa $M=(Q,\Sigma,\delta,r)$ where $r$ is the unique reject state.
$\Sigma = 2^{(I \cup O)}$ is the alphabet and $\delta: Q \times \Sigma \rightarrow Q$ with $r$ being the SINK state.
We assume that $\delta$ is encoded as multi terminal BDD as in MONA. Being MPNC it is assumed that 
\[
 \forall q \not=r. ~\forall i \in 2^I. ~ \exists o \in 2^O. \delta(q, i \cup o) \in Q-\{r\}
\]

\paragraph{Problem} Given MPNC $M$ and soft goals a sequence of literals $(l_1,l_2, \ldots, l_r)$ where $l_i= o$ or $l_i = \neg o$ for $o \in O$,
aim is to construct LODC $N$ by choosing exactly one of permitted outputs which maximizes the value of soft goal list 
(considered lexicographically ordered). Note that LODC satisfies the condition
\[
\begin{array}{l}
 \forall q \not=r. ~\forall i \in 2^I. ~ \\ 
  (\exists ! o \in 2^O. \delta_{lodc}(q, i \cup o) \in Q-\{r\} ~~\land \\
   \hspace*{1cm} (\forall o' \in 2^O. \delta_{mpnc}(1,i \cup o') \in Q-\{r\} ~\Rightarrow~ val(i \cup o') \leq val(i \cup o)) )
\end{array}
\]

{\bf We assume that in the bdd representation all $O$ occur after $I$}. The BDD node which is either terminal or labelled by output variable such
that all its ancestors are input labelled is called a {\bf frontier} node. See the example below.
\begin{figure}
\label{fig:mtbddrepresentation}
\begin{center}
\resizebox{7cm}{!}{
\includegraphics{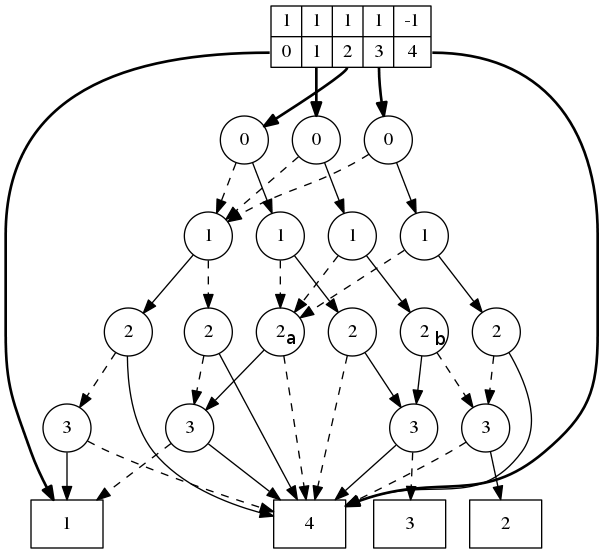}
}
\end{center}
\caption{BDD-representation of a Transition function.}
\end{figure}
In the Figure ~\ref{fig:mtbddrepresentation}, bdd nodes marked 0,1,2,3 correspond to variables $req1$, $req2$, $ack1$ and $ack2$. Nodes marked 2 are the frontier nodes.
Note that 4 is the reject state. For the leftmost frontier node marked (2), we can see that $ack1=true$ as well as $ack1=false, ack2=false$ lead to reject state 4 (these are infeasible paths), where as $ack1=false, ack2=true$ leads to good state 1. Hence $ack1=false, ack2=true$ is the unique feasible path.
Now consider the second rightmost frontier node marked (2). It has two feasible paths $P1=(ack1=true, ack2=false)$ going to target state 3 and  $P2=(ack1=false, ack2=true)$ going to target state 2. 

\paragraph{Step 1} Our algorithm works by assigning to each frontier node N  an output valuation, a reward value to each frontier node and
the target state.  Note that under given soft goal list, an optimal value can be  assigned to a frontier node purely by choosing an output path, 
from all feasible outputs (i.e. outputs which don't end in reject node $r$), one which maximizes the value of  the softgoal list.
This path ends in the target state.
\begin{example}
For example, in above figure for the second rightmost frontier node marked (2b) has
It has two feasible paths $P1=(ack1=true, ack2=false)$ going to target state 3 and  $P2=(ack1=false, ack2=true)$ going to target state 2. 
Given soft goal $(ack1,ack2)$, path $P1$ has higher lexicographic value and must be selected as optimal. Thus, the node can be marked with
$optoutput= (ack1=true, ack2=false)$, $optvalue=(1,0)$ and $optstate=3$.
\end{example}

\paragraph{Step 2} we can systematically enumerate all the paths to frontier nodes (mentioning only source state and the value of variables which 
occur on the path).  It is clear that for any two such paths originating from the same source state the value of at least one of the input variables 
is different, hence we get distinct cases.
(There can be identical input paths starting from different states. see example below.) 
For each such path, we set the output and next state to the output valuation and target state of the frontier node.
\begin{example}
 For the second rightmost frontier node marked (2-b), we have a unique input path $req1=true, req2=true$ which originates from state 1. 
 In conjunction with above above example, we get the scade transition
 \begin{verbatim}
   state=1 AND req1 AND req2 -> ack1=true, ack2=true, state=3
 \end{verbatim}
 For the third leftmost state marked (2-a) there are three input paths with $req1=true, req2=false$ originating in states either 1,2 or 3. 
\end{example}

The psuedocode of our algorithm is given as follows:
\begin{verbatim}
for each frontier node N 
  { 
    optoutput[N]:= bottom, optvalue[N] = -1, optstate:= bottom;
    for each output-path, PATH, to a non-rejecting state
       {
         output,value,endstatestate:= maxoutput(PATH,SOFTGOAL)
         if optvalue[N] < value
            {optoutput[N] := output;
             optvalue[N] := value;
             optstate[N] := state;
            }
        }
 }       
// the frontier node now has optimal output and its reward value
     
//Next we generate  LODC DFA below 
 
Initialize a new DFA with same states as MPNC and 
       with default transition to the unique reject state r.
       All other states are accept states.

for each each source state,
   for each input path, PATH, to a frontier node N
      {
         add Transition(PATH,optoutput[N],optstate[N]) to DFA
      }
 
LODC := Minimized(DFA)  // this gets rid of unreachable state  
\end{verbatim}

\clearpage
%\section{Case Study: N Cell Arbiter}     
%\label{EffSynCaseStudy}
\section{Case studies}
\label{section:casestudy}
In this section we give few sythesis case studies in our logic formalism.
We show the controller synthesis from the logic specification and
the effect of soft requirements on the sythesized controller.
We also compare the sythesized controllers based on our performance
measurement algorithm. Performance parameters therefore 
gives the quantitative matrics of betterness of one controller
over the other and can be used as the \emph{formal guarantees} 
for the sythesized controller with respect to the soft requirements.

\subsection{MinePump}
\label{section:minepumpexample}
In this section we present the detailed specification of a minepump controller. 
We first specify some useful generic properties which would 
used for requirement specification in case studies. 
\begin{itemize}
\item $lag(P,Q,n)$: 
%it is defined by Fig.~\ref{fig:lags2}. 
specifies that in any observation interval 
if $P$ holds continuously for $n+1$ cycles and persists 
then $Q$ holds from $(n+1)^{th}$ cycle onwards 
and persists till $P$ persists.
This specification is represented by following formula.
\[
%\begin{small}
\begin{array}{l}
\verb|[]([[P]] && slen>=n-1 => slen=n-1^[[Q]])|
\end{array}
%\end{small}
\]
    
\item $tracks(P,Q,n)$: in any observation interval if $P$ 
is continuously true for $n$ cycles then 
$Q$ persists as long as $P$ persists 
or for $n$ cycles whichever is shorter.
\[
%\begin{small}
\begin{array}{l}
\verb|{P} <=n= {Q}|
\end{array}
%\end{small}
\]
\item $sep(P, n)$:  
any interval which begins with a falling edge of $P$ and 
ends with a rising edge of $P$ then the length 
of the interval should be at least $n$ cycles.
\[
%\begin{small}
\begin{array}{l}
\verb|[]([P]^[!P]^<P> => slen > n)|
\end{array}
%\end{small}
\]
\item $ubound(P,n)$:  
in any observation interval $P$ can be continuously true 
for at most $n$ cycles.
\[
%\begin{small}
\begin{array}{l}
\verb|[]([[P]] => slen < n)|
\end{array}
%\end{small}
\]
\end{itemize}
%
%\begin{figure}[!h]
%\begin{minipage}{.49\textwidth}
%\centering
%\includegraphics[height=25mm, keepaspectratio]{fig/lags.png}
%\caption{$lags(P,Q,n)$.}
%\label{fig:lags2}
%\end{minipage}
%\begin{minipage}{.49\textwidth}
%\centering
%\includegraphics[height=25mm,keepaspectratio]{fig/tracks.png}
%\caption{$tracks(P,Q,n)$.}
%\label{fig:tracks}
%\end{minipage}
%%\end{figure}
%%\begin{figure}[!h]
%\begin{minipage}{.49\textwidth}
%\centering
%\includegraphics[height=20mm, keepaspectratio]{fig/separation.png}
%\caption{$sep(P,Q,n)$.}
%\label{fig:separation}
%\end{minipage}
%\begin{minipage}{.49\textwidth}
%\centering
%\includegraphics[height=20mm, keepaspectratio]{fig/ubound.png}
%\caption{$ubound(P,Q,n)$.}
%\label{fig:ubound}
%\end{minipage}
%\end{figure}

\subsubsection{Case Study Description:}
\label{subsection:minepump}
Imagine a minepump which keeps the water level in a mine 
under control for the safety of miners. 
The pump is driven by a controller which can switch it \emph{on} and \emph{off}. 
Mines are prone to methane leakage trapped underground 
which is highly flammable. So as a safety measure 
if a methane leakage is detected the controller is not allowed to 
switch on the pump under no circumstances. 

The controller has two input sensors - 
HH2O which becomes 1 when water level is high, 
and HCH4 which is 1 when there is a methane leakage; 
and can generate two output signals - 
ALARM which is set to 1 to sound/persist the alarm, 
and PUMPON which is set to 1 to switch on the pump. 
The objective of the controller is to \emph{safely} operate 
the pump and the alarm in such a way that 
the water level is never dangerous, 
indicated by the indicator variable DH2O, 
whenever certain assumptions hold. 
We have the following assumptions on the mine and the pump.

\begin{itemize}
\item[-] Sensor reliability assumption: $ppref(DH2O\Rightarrow HH2O)$ where \\
$ppref(D)=\neg((\neg D)\verb|^|ext)$. 
If HH2O is false then so is DH2O. 
\item[-] Water seepage assumptions: $tracks(HH2O, DH2O, w)$. 
The minimum no.~of cycles for water level 
to become dangerous once it becomes high is $w$. 
\item[-] Pump capacity assumption: $lags(PUMPON, !HH2O, epsilon)$. 
If pump is switched on for at least $epsilon+1$ cycles 
then water level will not be high after $epsilon$ cycles. 
\item[-] Methane release assumptions: $sep(HCH4,zeta)$ and $ubound(HCH4,$ $kappa)$. 
The minimum separation between the two leaks of methane is $zeta$ cycles 
and the methane leak cannot persist for more than $kappa$ cycles. 
\item[-] Initial condition assumption: 
$init(\textless !HH2O\textgreater\ \&\&\ \textless!HCH4\textgreater, slen=0)$. 
Initially neither the water level is high nor there is a methane leakage. 
\end{itemize}
The commitments are:
\begin{itemize}
\item[-] Alarm control: 
$lags(HH2O, ALARM, delta)$ and $lags(HCH4,$ \\ $ALARM, delta)$ and 
$lags(!HH2O\ \&\&\ !HCH4, !ALARM, delta)$. 
If the water level is dangerous 
then alarm will be high after $delta$ cycles and 
if there is a methane leakage then alarm will be high after $delta$ cycles. 
If neither the water level is dangerous nor there is a methane leakage 
then alarm should be off after $delta$  cycle. 
\item[-] Safety condition: $ppref(!DH2O\ \&\&\ (HCH4 \Rightarrow !PUMPON))$.
The water level should never become dangerous and 
whenever there is a methane leakage pump should be off. 
\end{itemize}

%The complete specification for minepump controller 
%in our XML specification format can be found in Appendix~\ref{appendix:minepump}. 
We can automatically synthesize a controller 
for the values say 
$w=8$, $epsilon=2$, $zeta=10$, $kappa=1$, and 
$delta=1$. The complete textual DCSynth specification of minepump is given is figure ~\ref{fig:minepumpsrc}. 
%The tool DCSynth outputs a SCADE or NuSMV controller meeting the specification. If the specification is not realizable we output an explanation.

\subsubsection{Soft goals and Performance measurement}
We could also synthesize the minepump controllers with each one of the following as a {\bf soft requirement} giving three different controllers. 
(Note that these requirements are mutually contradictory. So they are not given together.)
\begin{itemize}
\item \textbf{MPV1}: Keep the pump switched on whenever possible \textit{(specified by soft requirement $PUMPON$)}.
\item \textbf{MPV2}: Keep pump off if there is a methane leak in last 2 cycles otherwise switch on the pump \textit{(specified by soft requirement ($CH4_{Last2Cyc}$ $=>$ $!PUMPON$), where $CH4_{Last2Cyc}$ indicates that there was a methane leak within last 2 cycles)}.
\item \textbf{MPV3}: Keep pump off as much as possible i.~e.~delay the switching on the pump as much as possible \textit{(specified by soft requirement $!PUMPON$)}.
\end{itemize}
We have measured the performance  of the three synthesized controllers each,
 taking into account one soft requirement at a time.
The performance is measured in terms the maximum amount of time (in cycles) 
for which the water level can remain high  
without violating the assumptions  
and the detailed results are tabulated in Table.~\ref{tab:perfMeasure2}. 

\subsubsection{Simulation of Synthesized Minepump Controllers}
\begin{figure*}[!h]
%\begin{minipage}{.49\textwidth}
%\begin{center}
\centering
\includegraphics[width=\textwidth, keepaspectratio]{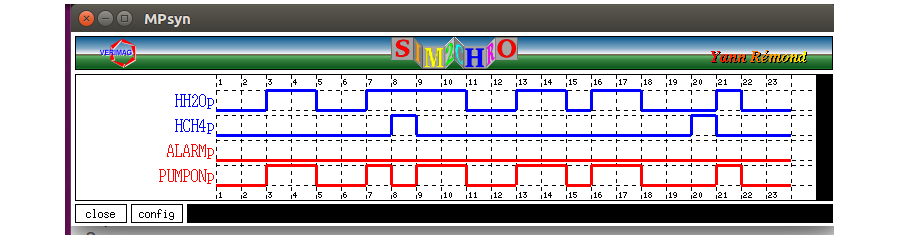}
%\end{center}
\caption{Simulation of Minepump Controller Variant MP\_V1}
\label{fig:pumptrue}
%\end{minipage}
%\end{figure}
%\begin{figure}[!h]
%\begin{center}
%\begin{minipage}{.49\textwidth}
%\begin{center}
\centering
\includegraphics[width=\textwidth, keepaspectratio]{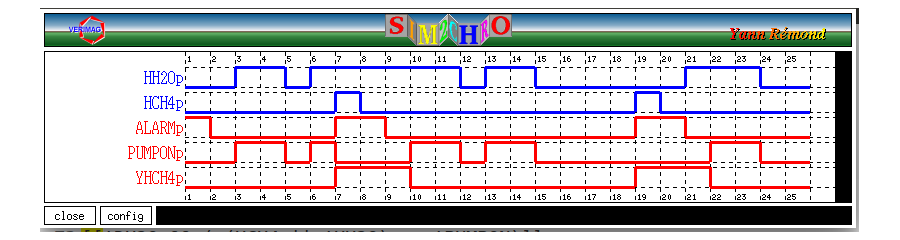}
%\end{center}
\caption{Simulation of Minepump Controller Variant MP\_V2}
\label{fig:pumpsoft}
%\end{minipage}
%\end{figure}
%
%\begin{figure}[!h]
%\begin{center}
%\begin{minipage}{.49\textwidth}
%\begin{center}
\centering
\includegraphics[width=\textwidth, keepaspectratio]{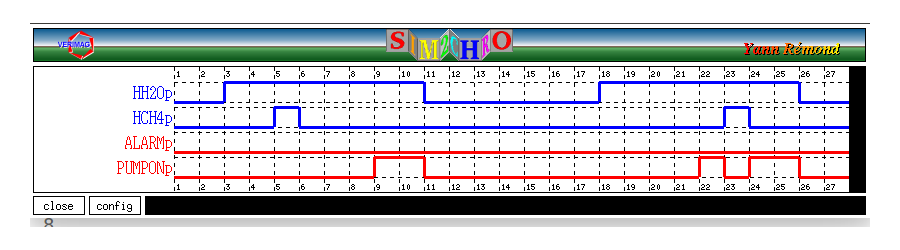}
%\end{center}
\caption{Simulation of Minepump Controller Variant MP\_V3}
\label{fig:pumpfalse}
%\end{minipage}
\end{figure*}

The controllers are encoded as Lustre specification and Lustre V4 tools are used 
for simulation. The example simulation for these three variants of Minepump
is shown in figures ~\ref{fig:pumptrue}, ~\ref{fig:pumpsoft} and ~\ref{fig:pumpfalse} respectively.

%We have tried to sythesize minepump controller by using tools Lily and Acacia+. Lily has timeout and Acacia+ could not determine the result (cf.~Table.~\ref{tab:comparision} last line). The minepump requirement in LTL were specified without any soft requirements as it is not supported by Lily and Acacia+. 

\section{Experimental results and Performance Evaluation}
\label{section:results}

This section is divided into two subsections. One deals with the performance of
the tools and its comparision with the other state of the art tools. 
The implementation of tool \dcsynth is based on the algorithms presented in 
Section ~\ref{subsection:algorithms}.  The second section deals with the performance
evaluation of synthesized controller. This allows the comparision of 
controllers produced for the same hard requirement with different
soft-requirements based on user defined performance measure specified
as maxlen of a QDDC formula.

\subsection{Tool Performance}
\label{subsection:toolperformance}

Table \ref{tab:comparision} shows how DCSynth fares against 
Acacia+, which is a leading tool for synthesis of controllers from temporal logic specification.
Acacia+ can handle LTL and PSL specification as well as quantitative synthesis  with mean payoff objectives. 
In contrast, our tool can only handle past time temporal properties but  it can handle {\em soft requirements} which other tools like Acacia+ cannot. 
Table \ref{tab:comparision} compares the performance of \DCSYNTH\/ against Acacia+ for examples with only hard safety requirements. 
It is noteworthy that controllers for complex specifications such as $MINEPUMP$
could be synthesized with \DCSYNTH. Table \ref{tab:comparisionopt} gives the results of synthesis using \DCSYNTH\/ for specifications which include
soft requirements.

In the table \ref{tab:comparision} the example $Arb^{tok}(n)$ indicates the n cell arbiter,
with a token circulating between them. Apart from the three invariant specifications (ARBINV) for the arbiter
given in section \ref{section:motivation} it specifies the requirement that if a cell has 
the request line true and it also has the token then it would surely get an acknowledgment.
The specification also says that the initially token is with cell 0, in next cycle
token is owned by cell 1, then 2 and so on till it reaches the last cell. 
At this time the token comes back to cell 0.
 
Similarly MP indicates the minepump example without soft requirement specification as given in section ~\ref{subsection:minepump}.
 
\begin {table}[!h]
\caption {Comparison of Synthesis in Acacia+ and \DCSYNTH}
\label{tab:comparision}
\begin{minipage}{\textwidth}
        \begin{center}
        \begin{tabular}
        %{|c|c|c|c|c|c|c|}
        {|p{0.2\linewidth} | p{0.2\linewidth}| p{0.2\linewidth}| p{0.2\linewidth}| p{0.2\linewidth}|} 
                \hline
                Problem &    \multicolumn{2}{|p{0.4\linewidth}|}{ Acacia+ }    & \multicolumn{2}{|p{0.4\linewidth}|}{ DCSynth }\\
                \hline
                & time (Sec) & Memory /States  & time (Sec) & Memory /States \\
                \hline  
%                $Arb^{hard}(2,2)$ & 0.37 & 28.2/ 5 & 0.01 & 4.6/ 2 \\
%                \hline    
%                $Arb^{hard}(3,3)$ & 1.67 & 60.7/ 13 &  0.02 & 4.6/ 8 \\
%                \hline    
                $Arb^{hard}(4,4)$ & 0.4 & 29.8/ 55 &  0.08 & 9.1/ 50 \\
                \hline    
                $Arb^{hard}(5,5)$ & 11.4 & 71.9/ 293 & 5.03 & 28.1/ 432 \\
                \hline    
                $Arb^{hard}(6,6)$ & TO\footnote{TO=timeout}  & - & ~ 80 & 1053.0/ 4802\\
                \hline
                $Arb^{hard}(7,7)$ & TO & - & - & MO\footnote{MO=memory out}\\
                \hline  
                $Arb^{tok}(7)$ & 9.65 & 39.1/ 57 & 0.3 & 7.3/ 7 \\
                \hline    
                $Arb^{tok}(8)$ & 46.44 & 77.9/ 73 &  2.2 & 16.2/ 8\\
                \hline    
                $Arb^{tok}(10)$ & NC\footnote{NC=synthesis inconclusive} & - &  152 & 82.0/ 10 \\
                \hline    
                $Arb^{tok}(12)$ & NC & - & TO & 255.0/ 12\\
                \hline  
%		Arbiter\_GR1\_2Cell & 0.3 & 17.7/ 3 & 0.37 & 28.0/ 4 & NE & - \\
%		\hline		
%		Arbiter\_GR1\_3Cell & 2.28 & 24.9/ 11 &  0.64 & 28.4/ 8 &  NE & - \\
%		\hline		
%		Arbiter\_GR1\_4Cell & 226.4 & 285.2/ 83  & 6.6 & 135.9/ 20 &  NE & - \\
%		\hline		
%		Arbiter\_GR1\_5Cell & TO & - & 1.5 & 41.2/ 173 & NE & -\\
%		\hline		
%		Arbiter\_GR1\_6Cell & - & - & 1153.4 & 330.8/ 1131 & NE & -\\
%		\hline
%		Arbiter\_GR1\_7Cell & - & - & TO & - & NE & -\\
%		\hline 
%		\hline
		MINEPUMP & NC & - & 0.06 & 50/ 32\\
		\hline
%		Minepump\_Soft\_PumpOff & NE & - & NE & - & 0.06 & 4.9/ 87 \\
%		\hline
%		Minepump\_Soft\_PumpOn & NE & - & NE & - & 0.06 & 50/ 32 \\
%		\hline
%		Minepump\_Soft\_MethanSafe & NE & - & NE & - & 0.13 & 62.8/ 43 \\
%		\hline
	\end{tabular}
\end{center}
\end{minipage}
\end{table}

\clearpage
\section{Minepump Source}

%As mentioned in \S\ref{subsection:dcvalid} the synthesized controller 
%is correct by construction meeting the minepump specification.

%\begin {table}[!h]
%\caption {Minepump Performance Measurement}
%\label{tab:softMP}
%\begin{center}
%	\begin{tabular}	{|c|c|c|c|}
%	\hline
%		Sr.No & Example & MAXLEN Formula & Response \\
%				& & Using MAXLEN & (In Cycle)\\
%	 \hline
%		1 & MP\_V1 & [[AssumptionOk \&\& HH2O]] & 4 \\
%	 \hline
%	    2 & MP\_V2 & [[AssumptionOk \&\& HH2O]] & 6 \\
%	 \hline
%	    3 & MP\_V3 & [[AssumptionOk \&\& HH2O]] & 8 \\
%	 \hline
%	\end{tabular}
%\end{center}
%\end{table}
%

%\subsection{Minepump Example Specification in QDDC}
%\label{subsec:minepumpspec}
\begin{figure}[!b]
\begin{small}
\framebox{\parbox[t][][t]{\columnwidth}{
$\begin{array}{l}
\mathrm{\textsf{BEGIN QDDCSYNTH MinePump}}\\
\quad\mathrm{\textsf{INTERFACESPEC}}\\
\quad\quad\mathrm{\textsf{$HH2Op$: INPUT}}\\
\quad\quad\mathrm{\textsf{$HCH4p$: INPUT}}\\
\quad\quad\mathrm{\textsf{$ALARMp$: OUTPUT MONITOR x}}\\
\quad\quad\mathrm{\textsf{$PUMPONp$: OUTPUT MONITOR x}}\\
\quad\quad\mathrm{\textsf{$YHCH4p$: OUTPUT MONITOR x;}}\\
\quad\mathrm{\textsf{SOFTREQS}}\\
\quad\quad\mathrm{\textsf{((!$YHCH4p$)$|$(!$PUMPONp$))$>>$($PUMPONp$) ;}}\\
\quad\mathrm{\textsf{AUXVARS}}\\
\quad\quad\mathrm{\textsf{$DH2O$}}\\
\quad\mathrm{\textsf{;}}\\
\quad\mathrm{\textsf{CONSTANTS}}\\
\quad\quad\mathrm{\textsf{-- delta response time of PUMP and ALARMS after trigger}}\\
\quad\mathrm{\textsf{delta = 1, w = 8, epsilon=2 , zeta=10, kappa=1}}\\
\quad\mathrm{\textsf{;}}\\
\quad\mathrm{\textsf{DEFINE}}\\
\quad\mathrm{\textsf{-- Alarm control}}\\
\quad\mathrm{\textsf{define alarm1(HH2O, DH2O, HCH4, ALARM, PUMPON) as  }}\\
\quad\quad\mathrm{\textsf{  {HH2O} $=delta=>$ {ALARM} ;}}\\
\quad\mathrm{\textsf{define alarm2(HH2O, DH2O, HCH4, ALARM, PUMPON) as}}\\
\quad\quad\mathrm{\textsf{  {HCH4} $=delta=>$ {ALARM} ;}}\\
\quad\mathrm{\textsf{define alarm3(HH2O, DH2O, HCH4, ALARM, PUMPON) as}}\\
\quad\quad\mathrm{\textsf{  { !HCH4 \&\& !HH2O } $=delta=>$ {!ALARM}   ;}}\\
\quad\mathrm{\textsf{-- Water seepage Assumptions}}\\
\quad\mathrm{\textsf{define water1(HH2O, DH2O, HCH4, ALARM, PUMPON) as }}\\
\quad\quad\mathrm{\textsf{   [] ( [[ DH2O $=>$ HH2O ]] ) ;}}\\
\quad\mathrm{\textsf{define water2(HH2O, DH2O, HCH4, ALARM, PUMPON) as}}\\
\quad\quad\mathrm{\textsf{   {HH2O} $<=w=$ {! DH2O}  ;}}\\
\quad\mathrm{\textsf{-- Pump capacity assumption}}\\
\quad\mathrm{\textsf{define pumpcap1(HH2O,DH2O,HCH4,ALARM,PUMPON) as}}\\
\quad\quad\mathrm{\textsf{  {PUMPON} $=epsilon=>$ {!HH2O} ;}}\\
\quad\mathrm{\textsf{--  Methane Release assumptions}}\\
\quad\mathrm{\textsf{define methane1(HH2O,DH2O,HCH4,ALARM,PUMPON) as}}\\
\quad\quad\mathrm{\textsf{   [] ( [HCH4]$\land$[!HCH4]$\land$$<$HCH4$>$ $=>$ slen $>$ zeta ) ;}}\\
\quad\mathrm{\textsf{define methane2(HH2O,DH2O,HCH4,ALARM,PUMPON) as}}\\
\quad\quad\mathrm{\textsf{   [] ( [[HCH4]] $=>$ slen $<$ kappa ) ;}}\\
\quad\mathrm{\textsf{-- Initial condition assumption}}\\
\quad\mathrm{\textsf{define initdry(HH2O, DH2O, HCH4, ALARM, PUMPON) as}}\\
\quad\quad\mathrm{\textsf{  $<$!HH2O$>$ $\land$ true ;}}\\
\quad\mathrm{\textsf{-- safety condition }}\\
\quad\mathrm{\textsf{define safe(HH2O, DH2O, HCH4, ALARM, PUMPON) as}}\\
\quad\quad\mathrm{\textsf{[[!DH2O \&\& ( (HCH4 $||$ !HH2O) $=>$ !PUMPON)]];}}\\
\end{array}$}
}
\end{small}
\end{figure}
\begin{figure}[!b]
\begin{small}
\framebox{\parbox[t][][t]{\columnwidth}{
$\begin{array}{l}
\quad\mathrm{\textsf{define plant(HH2O, DH2O, HCH4, ALARM, PUMPON) as}}\\
\quad\quad\mathrm{\textsf{initdry(HH2O, DH2O, HCH4, ALARM, PUMPON) \&\&}}\\
\quad\quad\mathrm{\textsf{water1(HH2O, DH2O, HCH4, ALARM, PUMPON) \&\&}}\\
\quad\quad\mathrm{\textsf{water2(HH2O, DH2O, HCH4, ALARM, PUMPON) \&\&}}\\
\quad\quad\mathrm{\textsf{pumpcap1(HH2O, DH2O, HCH4, ALARM, PUMPON) \&\&}}\\
\quad\quad\mathrm{\textsf{methane1(HH2O, DH2O, HCH4, ALARM, PUMPON) \&\&}}\\
\quad\quad\mathrm{\textsf{methane2(HH2O, DH2O, HCH4, ALARM, PUMPON);}}\\
\quad\mathrm{\textsf{define req(HH2O, DH2O, HCH4, ALARM, PUMPON) as}}\\
\quad\mathrm{\textsf{infer MPsyn as}}\\
\quad\mathrm{\textsf{INDICATORS }}\\
\quad\quad\mathrm{\textsf{ YHCH4p : (slen=2 \&\& $<><$HCH4p$>$)}}\\
\quad\mathrm{\textsf{;}}\\
\quad\mathrm{\textsf{ASSUME}}\\ 
\quad\quad\mathrm{\textsf{plant(HH2Op, DH2O, HCH4p, ALARMp, PUMPONp)}}\\
\quad\mathrm{\textsf{;}}\\
\quad\mathrm{\textsf{REQUIRES}}\\ 
\quad\quad\mathrm{\textsf{req(HH2Op, DH2O, HCH4p, ALARMp, PUMPONp)}}\\
\quad\mathrm{\textsf{;}}\\
\quad\mathrm{\textsf{SYNTHESIZE}}\\
\quad\mathrm{\textsf{SynthG MPsyn}}\\
\quad\mathrm{\textsf{END QDDCSYNTH}}\\
\end{array}$}
}
\end{small}
\caption{DCSynth spec for Minepump.}
\label{fig:minepumpsrc}
\end{figure}

\clearpage
\section{Textual Specification of Arbiter Case Study}
\label{section:arbiterqddc}

The DCSynth specification of the 4 cell arbiter is shown in figure ~\ref{fig:arbitersrc}.
It can be easily generalized to n-cells.

%\subsection{Source Code}
%\label{•}label{arbsrc}
\begin{figure}[!h]
\begin{small}
\framebox{\parbox[t][151mm]{\columnwidth}{
$\begin{array}{l}
\mathrm{\textsf{BEGIN QDDCSYNTH arbiter-4-cell}}\\
\quad\mathrm{\textsf{INTERFACESPEC}}\\
\quad\mathrm{\textsf{$req1$: INPUT}}\\
\quad\mathrm{\textsf{$req2$: INPUT}}\\
\quad\mathrm{\textsf{$req3$: INPUT}}\\
\quad\mathrm{\textsf{$req4$: INPUT}}\\
\quad\mathrm{\textsf{$ack1$: OUTPUT MONITOR x}}\\
\quad\mathrm{\textsf{$ack2$: OUTPUT MONITOR x}}\\
\quad\mathrm{\textsf{$ack3$: OUTPUT MONITOR x}}\\
\quad\mathrm{\textsf{$ack4$: OUTPUT MONITOR x}}\\
\quad\mathrm{\textsf{$sr1$: OUTPUT MONITOR x}}\\
\quad\mathrm{\textsf{$sr2$: OUTPUT MONITOR x}}\\
\quad\mathrm{\textsf{$sr3$: OUTPUT MONITOR x}}\\
\quad\mathrm{\textsf{$sr4$: OUTPUT MONITOR x}}\\
\quad\mathrm{\textsf{ ;}}\\
\quad\mathrm{\textsf{SOFTREQS}}\\
\quad\mathrm{\textsf{($sr4$)$>>$($sr3$)$>>$($sr2$)$>>$($sr1$)}}\\
\quad\mathrm{\textsf{;}}\\
\quad\mathrm{\textsf{CONSTANTS}}\\
\quad\mathrm{\textsf{;}}\\
\quad\mathrm{\textsf{AUXVARS}}\\
\quad\mathrm{\textsf{;}}\\
\quad\mathrm{\textsf{DEFINE}}\\
\quad\mathrm{\textsf{-- Acknowledgments should be exclusive}}\\
\quad\mathrm{\textsf{define exclusion() as}}\\
\quad\mathrm{\textsf{[[((ack1 $=>$ !((ack2) $||$ (ack3) $||$ (ack4))) \&\&}}\\
\quad\mathrm{\textsf{(ack2 $=>$ !((ack1) $||$ (ack3) $||$ (ack4))) \&\& }}\\
\quad\mathrm{\textsf{(ack3 $=>$ !((ack1) $||$ (ack2) $||$ (ack4))) \&\& }}\\
\quad\mathrm{\textsf{(ack4 $=>$ !((ack1) $||$ (ack2) $||$ (ack3))))]];}}\\
\quad\mathrm{\textsf{-- No lost cycle}}\\
\quad\mathrm{\textsf{define noloss() as}}\\
\quad\mathrm{\textsf{[[((req1) $||$ (req2) $||$ (req3) $||$ (req4))}}\\
\quad\mathrm{\textsf{ $=>$ ((ack1) $||$ (ack2) $||$ (ack3) $||$ (ack4))]];}}\\
\quad\mathrm{\textsf{-- Ack should be granted if there is a request}}\\
\quad\mathrm{\textsf{define nospuriousack(ack, req) as}}\\
\quad\mathrm{\textsf{[[ack $=>$ req]];}}\\
\quad\mathrm{\textsf{-- 2 cycle response property}}\\
\quad\mathrm{\textsf{define response(req, ack) as}}\\
\quad\mathrm{\textsf{[ ](($[[req]]$ \&\& (slen=k-1)) $=>$ $<><ack>$);}}\\%}}\\
\end{array}$}}
\end{small}
\end{figure}
\begin{figure}[!h]
\begin{small}
\framebox{\parbox[t][]{\columnwidth}{
$\begin{array}{l}
\quad\mathrm{\textsf{infer arbiter4cell as}}\\
\quad\mathrm{\textsf{INDICATORS}}\\
\quad\mathrm{\textsf{sr1 : (response(req1, ack1))}}\\ 
\quad\mathrm{\textsf{sr2 : (response(req2, ack2))}}\\ 
\quad\mathrm{\textsf{sr3 : (response(req3, ack3))}}\\
\quad\mathrm{\textsf{sr4 : (response(req4, ack4))}}\\ 
\quad\mathrm{\textsf{;}}\\
\quad\mathrm{\textsf{ASSUME}}\\
\quad\mathrm{\textsf{;}}\\
\quad\mathrm{\textsf{REQUIRES}}\\
\quad\mathrm{\textsf{  exclusion()}}\\
\quad\mathrm{\textsf{  noloss()}}\\
\quad\mathrm{\textsf{  nospuriousack(ack1, req1)}}\\
\quad\mathrm{\textsf{  nospuriousack(ack2, req2)}}\\
\quad\mathrm{\textsf{  nospuriousack(ack3, req3)}}\\
\quad\mathrm{\textsf{  nospuriousack(ack4, req4)}}\\
\quad\mathrm{\textsf{ ; }}\\
\quad\mathrm{\textsf{SYNTHESIZE}}\\
\quad\mathrm{\textsf{ SynthG arbiter4cell}}\\
\quad\mathrm{\textsf{END QDDCSYNTH}}\\
\end{array}$}}
\end{small}
\caption{DCSynth spec for 4 Cell Arbiter.}
\label{fig:arbitersrc}
\end{figure}

%amol
%amol
%amol
\end{document}